\documentclass[prd,aps,preprintnumbers]{revtex4}
\usepackage{epsfig}
\usepackage{amsmath}
\usepackage{hyperref}

\def\asbar{{\bar {\ensuremath{\alpha}}}_{s}}
\def\as{\ensuremath{\alpha_{s}}}

\def\MSbar{$\overline{{\rm MS}}$}
\def \ep{\epsilon}
\def\s{\sigma}
\def\xij{x_{ij}}
\def\yij{y_{ij}}

\def\bea {\begin{eqnarray}}
\def\eea {\end{eqnarray}}

\def\be {\begin{equation}}
\def\ee {\end{equation}}

\begin{document}

\preprint{PITHA 09/17}
\preprint{ITP-SB-09-20}

\renewcommand{\thefigure}{\arabic{figure}}

\title{Threshold Resummation for Top-Pair Hadroproduction\\ to Next-to-Next-to-Leading Log}

\author{Michal Czakon}
\affiliation{Institut f\"ur Theoretische Physik E, RWTH Aachen University,
D-52056 Aachen, Germany}
\author{Alexander Mitov, George Sterman}
\affiliation{C.N.\ Yang Institute for Theoretical Physics,
Stony Brook University, Stony Brook, New York 11794--3840, USA}

\date{\today}

\begin{abstract}
We derive the threshold-resummed total cross section for heavy quark
production in hadronic collisions accurate to next-to-next-to-leading logarithm, employing
recent advances on soft anomalous dimension
matrices for massive pair production
in the relevant kinematic limit.
We also derive the relation between
heavy quark threshold resummations
 for fixed pair kinematics and the inclusive cross section.
 As a check of our results, we have verified
 that they reproduce all poles of the color-averaged
 $q\bar{q}\rightarrow t\bar{t}$ amplitudes at two
 loops, noting that the latter are insensitive to
 the color-antisymmetric terms of the soft anomalous dimension.
 \end{abstract}

\maketitle


\section{Introduction}

Heavy quark production has a long and continuing
history of phenomenological interest. Top-pair production
is among the most important standard model
processes in the context of searches for new physics.
It is also among the most challenging in computation.
Indeed over 20 years passed between the
landmark first NLO calculations of heavy quark production
\cite{QQbarfixed} and the derivation \cite{Czakon:2008ii}
of analytic expressions for the inclusive
cross section.

As in many hard hadronic processes, higher order
perturbative calculations contain logarithmic enhancements,
associated with the approach to partonic threshold, that is,
configurations where the initial-state partons have just
enough energy to produce the observed final state.
Threshold resummation \cite{dyresum} organizes such logarithms,
in a manner we review below.
The current state of the art for heavy quark production is next-to-leading log
(NLL) resummation matched onto exact
next-to-leading order (NLO) results.
In this paper, we study the extension of this
formalism to  next-to-next-to-leading log
(NNLL) resummation and beyond in the production of heavy quark-antiquark pairs at
hadron colliders.   In this context, we derive the NNLL inclusive cross section,
generalizing the NLL results of \cite{Bonciani:1998vc}, starting from
the resummation formalism at fixed parton kinematics
described in \cite{Kidonakis:1997gm}.
The numerical impact of
our present work will be detailed elsewhere.

While our considerations are reasonably general and can be applied
to a larger class of processes, we choose to present our results and
their derivation for heavy quark production because recent advances
\cite{Mitov:2009sv,Kidonakis:2009ev,Becher:2009kw} make it possible
for us to determine the essentially two-loop inputs necessary for
explicit NNLL resummation in this case.    These inputs are the
so-called soft anomalous dimension matrices for heavy quark pair
production \cite{Kidonakis:1997gm} which we exhibit below in the
relevant kinematic configuration for the total cross section.
Drawing from the formalism of Ref.\ \cite{Kidonakis:1997gm}, we will
perform a single one-loop calculation, corresponding to a boundary
condition in the evolution of soft gluon emission, which is
necessary for the complete  NNLL result. As noted at the level of
NLL, for the inclusive cross section, resummation can be carried out
separately for pair production in the $s$-channel color singlet and
octet states, without color mixing.   We will find the same
structure at NNLL, and  present our final results in a form that
follows the NLL formalism of Ref.\ \cite{Bonciani:1998vc}.

We begin with a review of threshold resummation for
semi-inclusive cross sections at fixed kinematics for the
partonic scattering process, in the formalism developed
in Ref.\ \cite{Kidonakis:1997gm}, and applicable
in principle to any logarithmic approximation.   This formalism
relies on the factorization
of color-diagonal ``jet" functions associated with the external
energetic and/or massive partons
that take part in the scattering at short distances and a ``soft function"
describing the exchange of low-energy
quanta between these particles.   We identify in particular,
a scheme to resolve the ambiguity between the jet and soft functions,
based on the singularities of form factors in dimensional regularization.
We go on in Section \ref{sec:reduction} to show how an expression for
the threshold-resummed inclusive cross section may be derived
from the resummation at fixed kinematics.   With this result in hand,
we determine two-loop soft anomalous dimension matrices in Sec.\ \ref{sec:gamma2},
which we need to determine the soft functions at NNLL.
We describe a very nontrivial check of these results, which
fully reproduce the two-loop pole structure of heavy quark pair production
in light quark annihilation.
In Sec.\ \ref{sec:NNLL-res-crossection}, we assemble the remaining ingredients in
the NNLL resummation, including the one-loop boundary
condition mentioned above.
We conclude with a summary and a few comments
on prospects for future work.

\section{Threshold Resummation at Fixed Kinematics}
\label{sec:soft-gluon-res}

In this section, we review the threshold resummation formalism
of Ref.\ \cite{Kidonakis:1997gm}, which is adapted to
semi-inclusive reactions
characterized by fixed partonic scattering kinematics, as in
for example,
\bea
f_1(p_1)+f_2(p_2) \rightarrow f_a(p_a) +f_b(p_b)\, ,
\label{2to2}
\eea
where $f_i(p_i)$ denotes a parton of flavor $f_i$
and momentum $p_i$.    We have shown a $2\rightarrow 2$
process, but final states with more than two particles are
also possible, so long as all invariants $p_i\cdot p_j$ are large.
The formalism we sketch in this section applies
to processes involving light quarks and gluons,
and also to the production of heavy quarks.
In the latter case, we can also study the inclusive cross section,
for which threshold resummation has been
developed from a related point of view \cite{Bonciani:1998vc}.   In Section \ref{sec:reduction}
we will derive resummed inclusive cross sections
for heavy quark production from their semi-inclusive forms.

\subsection{Factorization near partonic threshold}

Our starting point for the resummation of observables involving
initial and/or final state hadrons is the formalism of
Ref.~\cite{Kidonakis:1997gm}. To be specific, we restrict our
discussion to the $2\rightarrow 2$ processes of Eq.\ (\ref{2to2}),
although many of our considerations can be directly generalized. For
the production of a pair of particles with mass $m$, the kinematics
can be described by the invariant mass $M$ and rapidity $y$ of the partonic
final state and the pair center-of-mass rapidity difference
$\hat\eta$.   Assuming that $m\gg \Lambda_{\rm QCD}$, this cross
section can be written in standard factorized form as \bea
M^4\frac{d\sigma_{h_1h_2\rightarrow Q\bar{Q}}}{dM^2 dy d\hat\eta}
&=& \sum_f \int_\tau^1 dz \int \frac{dx_a}{x_a}\frac{dx_b}{x_b}\
\phi_{f/h_1}(x_a,\mu^2)\, \phi_{{\bar f}/h_2}(x_b,\mu^2)
\nonumber\\
&\ & \hspace{5mm} \times\ \delta\left( z -\frac{\tau}{x_ax_b}\right)\, \delta\left( y - \frac{1}{2}\ln\frac{x_a}{x_b}\right)
\nonumber\\
&\ &
\hspace{5mm} \times\ \omega_{f\bar{f}\rightarrow Q\bar{Q}}\left(z,\hat{\eta},\frac{M^2}{\mu^2},\frac{m^2}{\mu^2},\as(\mu^2)\right)\, ,
\label{eq:cofact}
\eea
where we have normalized the cross section so that all quantities are dimensionless.
The purpose of threshold resummation is
to organize plus distributions in the variable
\bea
z = \frac{\tau}{x_ax_b} = \frac{M^2}{x_ax_bS}\, ,
\eea
with $x_a$ and $x_b$ the usual fractional momenta.
 Partonic threshold
is defined as the limit $z\rightarrow 1$, at which the
incoming partons provide just enough energy to
produce the observed final state.   The mismatch
between real gluon emission and virtual corrections
gives rise to singular distributions at $z=1$.
These distributions appear in the $n$th order expansion of
the perturbative function $\omega_{f\bar{f}\rightarrow Q\bar{Q}}$
up to the level of $\as^n[\ln^{2n-1}(1-z)/(1-z)]_+$.

In Ref.\ \cite{Kidonakis:1997gm}, it was observed that as $z\rightarrow 1$,
 partonic cross sections can be
factorized into a set of universal factors associated with
the incoming and outgoing partons of the underlying
process, along with process-dependent factors
that describe the coherent interactions of those
partons, at short and long distances.   The resummed
dependence in $1-z$ is conveniently generated by taking
Mellin moments with respect to $z$, schematically,
\begin{eqnarray}
\sigma(N) &=&
\int_0^1 dz z^{N-1} \sigma(z)
\nonumber\\
&=& \int_0^1 dz e^{-(N-1)(1-z)} \sigma(z) + {\cal O}(1/N)\, .
\end{eqnarray}
For example, in the inclusive
Drell-Yan process, the corresponding kinematical variable is
$z=Q^2/s$, where $s$ is the partonic c.m. energy \cite{dyresum}.
For resummation of
the total inclusive cross section of  heavy quark pair production at hadron
colliders the corresponding variable is $z=4m^2/s$ where $m$ is the
mass of the top quark. Another example relevant for this paper is
the invariant mass $M_{t{\bar t}}^2=(p_t+p_{\bar t})^2$ distribution
of a top quark-antiquark pair; the relevant partonic variable is
$z=M_{t{\bar t}}^2/s$.
In any case we assume that the partonic
variable is defined such that threshold kinematics is attained in
the limit $z\to 1$. In moment space this corresponds to the limit
$N\to\infty$.
The analysis of Ref.\ \cite{Kidonakis:1997gm}
exploits factorization near threshold, according to which the cross section can be
written as a convolution in an appropriate momentum component
of the soft radiation associated with a set of functions \cite{Contopanagos:1996nh}.
In threshold resummation for hadronic collisions, this component is
the energy, $E_i^*$, of each final-state particle in the center-of-mass frame
of the hard collision.
That is,  for any threshold resummation at hadronic collisions, we can identify
\begin{eqnarray}
1-z = \sum_{{\rm particles}\ i} \frac{2E^*_i}{\sqrt{s}} \, ,
\end{eqnarray}
where the partonic variable $s \equiv x_ax_bS$ equals $M^2$ at threshold,
with $M$ the invariant mass of the observed pair of heavy particles. The cross
section then factorizes into simple products in the corresponding
moment space. Dependence on the moment variable enters only through
the transform, and is therefore always in the form $N/M$, up to
corrections that decrease as powers of $N$.

As a result of this analysis, the partonic cross section takes a
factorized form in moment space, which we can represent as
\bea
\label{eq:observable}
\omega_P\left(N,\hat{\eta},\frac{M^2}{\mu^2},\frac{m^2}{\mu^2},\as(\mu^2)\right)\
 &=& J_1(N,\as(\mu^2)) \dots
J_k(N,M/\mu,m/\mu,\as(\mu^2))
\nonumber\\
&\ & \hspace{3mm} \times\ {\rm Tr}
\left[ {\bf H}^P\left(\frac{M^2}{\mu^2},\frac{m^2}{\mu^2},\hat{\eta},\as(\mu^2)\right)
{\bf S}^P\left(\frac{N^2\mu^2}{M^2},\frac{M^2}{m^2},\hat{\eta},\as(\mu^2)\right)\right] +
{\cal O}(1/N) \, ,
\eea
where the label $P$ refers to a particular partonic process, for
example $q\bar{q}\rightarrow t\bar{t}$, with $q$ a light flavor. The
Mellin moment $N$ is conjugate to the kinematical variable $z$. As
shown, the various functions appearing in Eq.~(\ref{eq:observable})
depend on other kinematical variables and masses as well as the
factorization and renormalization scales. These functions depend on
the specific process. Below, we will give them more explicitly in
the specific examples considered here. We will refer to the factors
$J_i$ appearing in Eq.~(\ref{eq:observable}) as the {\it jet
functions} for the underlying process. They are color diagonal
functions that describe the factorized dynamics of initial and/or
final state hard partons, whether massive or massless,
and as such are independent of the details
of the hard subprocess. Jet functions for initial-state partons
absorb the collinear subtractions necessary to define the hard
scattering function $\omega$ in Eq.\ (\ref{eq:observable}), so that
they are infrared safe. Jet functions for final-state partons are
automatically infrared safe for the differential and inclusive cross
sections that we discuss here. The formalism can be extended as well
to a variety of jet observables and to single-hadron cross sections.
The number $k$ of such functions in Eq.~(\ref{eq:observable})
corresponds to the number of hard colored partons in the process
being considered.

The functions ${\bf H}$ and ${\bf S}$ appearing in
Eq.~(\ref{eq:observable}) are known as {\it hard} and {\it soft}
functions, respectively. They are both matrices in the space of
tensors that describe the exchange of color at short distances
\cite{Kidonakis:1997gm}.    Examples for quark-antiquark scattering
are color singlet or octet in the $s$- or $t$-channel. We will
denote these tensors in boldface, and their product is traced over
the combinations of color tensors in the amplitude and its complex
conjugate. In the limit $N\to\infty$ the hard function ${\bf H}$ is
free of logarithmic dependence on $N$; it is obtained from a
dedicated, process-specific calculation.

\subsection{Moment-dependence and the soft anomalous dimension matrix}
\label{sec:soft-matrix}

The soft function ${\bf S}$ contains terms due to wide-angle soft
emissions and thus contributes a single power of $\ln(N)$ per loop.
It is also process dependent, and in the general case is dependent
on the four-velocities $\{\beta_i\}$ of the partons that take part
in the hard scattering. For processes involving four or more colored
hard partons it is a matrix in the space of color tensors. Assuming
fixed-angle scattering, the soft function depends on the scalar
products of these velocities, in addition to a single overall scale,
which we will take to be $M$, the invariant mass of the pair for the
case of heavy quark production.  For a massive quark of velocity
$\beta_q$, we shall set $\beta_q^2=m_q^2/M^2$, and for most of this
discussion, treat this ratio as a number of order unity.

As noted above, all $N$-dependence is of the form $N/M$.   As a
result, in the dimensionless soft function, $N$-dependence appears
only in the combination $M/(N\mu)$.   In
Ref.~\cite{Kidonakis:1997gm}, it was shown that the $N$-dependence
of the soft function ${\bf S}(N,\dots)$ entering the cross section
Eq.~(\ref{eq:observable}) can be made explicit in terms of a ``soft
anomalous dimension matrix", ${\bf \Gamma}_S$. Making the natural
choice, $\mu=M$, we have
\begin{eqnarray}
\label{eq:S-observ} {\bf S}\left (\frac{N^2\mu^2}{M^2},\beta_i\cdot\beta_j,\as(\mu^2)\right)\, \Bigg |_{\mu=M}
 &=& {\cal {\overline
P}} \exp\Bigg\{-\ \int_{M/\bar{N}}^M \frac{d\mu'}{\mu'}
{\bf\Gamma}_S^\dag\left(\beta_i\cdot\beta_j,\as\left(\mu'{}^2\right) \right)
\Bigg\} \nonumber\\
&\ & \times {\bf S}\left(1,\beta_i\cdot\beta_j,\as\left(M^2/\bar{N}^2\right)\right)
\nonumber\\
&\ & \times {\cal P} \exp\Bigg\{-\ \int_{M/\bar{N}}^M \frac{d\mu'}{\mu'}
{\bf\Gamma}_S\left(\beta_i\cdot\beta_j,\as\left(\mu'{}^2\right) \right) \Bigg\}
\nonumber\\
&=& {\cal {\overline
P}} \exp\Bigg\{\int_0^1 dx{x^{N-1}-1\over 1-x} ~
{\bf\Gamma}_S^\dag\left(\beta_i\cdot\beta_j,\as\left((1-x)^2M^2\right) \right)
\Bigg\} \nonumber\\
&\ & \times {\bf S}\left(1,\beta_i\cdot\beta_j,\as\left(M^2/N^2\right)\right)
\nonumber\\
&\ & \times {\cal P} \exp\Bigg\{\int_0^1 dx{x^{N-1}-1\over 1-x} ~
{\bf\Gamma}_S\left(\beta_i\cdot\beta_j,\as\left((1-x)^2M^2\right) \right) \Bigg\}
\, ,
\end{eqnarray}
where the second expression is accurate to next-to-next-to leading
logarithms (i.e. terms $\sim \alpha_s^n\ln^{n-1}N$ in the cross
section) for $\bar{N}=Ne^{\gamma_E}$, with $\gamma_E$ the Euler
constant.   Throughout this paper $\as =
\as(\mu^2)$ is the standard \MSbar\ coupling evolving with $N_L$
light flavors. Decoupling of the heavy flavor will simplify our
results significantly. The relation between the bare $\as^b$ and
renormalized couplings reads
\begin{eqnarray}
\label{eq:alpha-s-renorm} \as^{\rm{b}} S_\epsilon \: = \: \as(\mu^2)
\biggl[
   1
   - {\beta_0 \over 4\epsilon}  {\as(\mu^2) \over \pi}
   + {\cal O}(\as^2)
  \biggr]
\, ,
\end{eqnarray}
where $S_\ep=(4\pi)^\ep \exp(-\ep \gamma_{\rm E})$ and $\beta_0 =
(11/3)C_A - (4/3)T_FN_L$. The color factors in an
${\rm{SU}}(N)$-gauge theory are $C_A = N$, $C_F = (N^2-1)/(2N)$ and
$T_F = 1/2$.

The structure of Eq.~(\ref{eq:S-observ}) follows from the
renormalization group equation satisfied by the soft function ${\bf
S}(N^2\mu^2/M^2,\dots)$, where ${\bf \Gamma}_S$ plays the role of a
matrix of anomalous dimensions \cite{Kidonakis:1997gm}. The function
${\bf S}(1,\dots)$ plays the role of a boundary condition, which is
chosen to be the soft function at unit $N$, that is, with unit
weight. In general, this factor contributes a single  $\ln(N)$
starting from two loops, which is due, however, entirely to the
presence of $N$ in the scale of the running coupling in its one-loop
expression. To determine this contribution one need only calculate
the soft function in Eq.~(\ref{eq:observable}) through one loop.

At $N=1$, the computation of the soft function is given by
a total eikonal cross section, subtracted for eikonal jet functions
to eliminate collinear enhancements \cite{Kidonakis:1997gm}.
In the formalism of Ref.\ \cite{Kidonakis:1997gm},
virtual corrections are pure counterterms, because the corresponding
eikonal diagrams are scaleless and vanish in dimensional regularization.
In the full soft function, however,
 the hard scale sets a maximum total
energy for the soft function at $N=1$, and the corresponding
integrals are not scaleless.   Their infrared poles are cancelled
by the virtual diagrams, but finite corrections may remain.

In summary, the soft function ${\bf S}$ at $N=1$ takes the form
\begin{equation}
\label{eq:S-tilde} {\bf S}
\left(1,\beta_i\cdot\beta_j,\as\left(M^2/N^2\right)\right) = {\bf S}^{(0)}
+ \frac{\as\left(M^2/N^2\right)}{\pi}\ {\bf S}^{(1)}\left(1,\beta_i\cdot\beta_j\right) + \dots\, ,
\end{equation}
where ${\bf S}^{(0)} $ is a constant diagonal matrix independent of
the coupling and ${\bf S}^{(1)}\left(1,\beta_i\cdot\beta_j\right)$
is free of dependence on $N$, but can depend on the eikonal
velocities that define the soft function. Explicit expressions for
${\bf S}^{(0)} $ relevant to heavy quark production can be found in
\cite{Kidonakis:2001nj}. We will give the one-loop correction below,
after specifying a scheme that defines the soft function unambiguously.
At this stage, we note that to compute the soft function fully at
next-to-next to leading logarithm it is necessary to compute the
two-loop anomalous dimension matrix and the one-loop soft function.

\subsection{The form factor scheme}
\label{schemesec}

The soft function is not unique, but is ambiguous at the level of
single logarithmic contributions that can be absorbed into the jet
functions. These ambiguities, must be proportional to the unit
matrix in the color exchange space (since the jet functions are
diagonal in color). To resolve this ambiguity one has to specify a
prescription for the definition of the anomalous dimension matrix
${\bf \Gamma}_S$, which we discuss next.

A fundamental observation of Ref.~\cite{Kidonakis:1997gm} is that
the matrix ${\bf \Gamma}_S$ appearing in Eq.~(\ref{eq:S-observ}) can
be extracted from the corresponding (virtual) amplitude for the
process under consideration. To review how this can be done, we
first observe  that any on-shell, renormalized scattering amplitude
at fixed angles can be factorized as follows \cite{Sen:1982bt}:
\begin{equation}
\label{eq:amplitude} M_I(\ep,\dots) = j_1(\ep,\dots) \dots
j_s(\ep,\dots)\cdot h_J(\dots) \cdot s_{IJ}(\ep,\dots) \, .
\end{equation}
Here $I,J$ are indices that label color exchange tensors; in
particular, they indicate that the amplitude $M$ can be thought of
as a vector in the space of color representations \cite{catani96}.
In order to emphasize the similarity between the objects appearing
in Eqs.~(\ref{eq:observable}) and (\ref{eq:amplitude}) we have used
the same letters to denote jet, soft and hard functions. It should
be stressed, however, that these are not the same objects.   In
particular, the moment $N$ does not appear in an amplitude. To
distinguish clearly  between the objects appearing in the
cross section and in the amplitude we use lower case letters, and
explicitly show the dependence on the infrared regulator $\ep$
where $d=4-2\ep$.

For both massive and massless external partons the amplitude soft
function $s_{IJ}(\ep,\dots)$ appearing in Eq.~(\ref{eq:amplitude})
is fully determined by its (matrix) anomalous dimension
$\Gamma_{IJ}$. It can be computed order by order in perturbation
theory, as a series in the coupling $\as(\mu^2)$. The properties of
$\Gamma_{IJ}$ have been studied extensively in the massless
\cite{Aybat,Becher:2009cu,Gardi:2009qi,Becher:2009qa} and in the
massive \cite{Mitov:2009sv,Kidonakis:2009ev,Becher:2009kw} cases. To
unambiguously fix the soft function in the massless case, a natural
and convenient scheme was proposed in Ref.~\cite{TYS}. There, the
jet functions for each external parton were identified with the
square root of the massless on-shell space-like form factor for the
corresponding parton. We will assume this prescription, which we
call {\it the form factor scheme} by default from now on. In this
scheme, the jet functions for massless particles are series in the
coupling $\as(\mu^2)$ with coefficients that are $\ep$-dependent
numbers. A natural extension \cite{Mitov:2006xs} in the massive case
is to identify the jet function with the small-mass limit of the
corresponding massive space-like QCD form factor. Therefore, in the
massive case the jet function contains also powers of
$\ln(\mu^2/m^2)$, where $m$ is the pole mass of the heavy quark.

In the form factor scheme,
we can derive an explicit expression for ${\bf s}(\ep,\dots)$
in terms of the matrix soft anomalous dimension,
\begin{equation}
\label{eq:S-function-amplitude} {\bf s}(\ep,\dots)  = {\cal P}
\exp\Bigg\{ - \int_0^1 {dx\over 1-x} ~
{\bf\Gamma}_S\left(\asbar\left[(1-x)^2Q^2\right]\right) \Bigg\}\, ,
\end{equation}
where ${\bf\Gamma}_S = (a/\pi){\bf \Gamma}_S^{(1)} + (a/\pi)^2{\bf
\Gamma}_S^{(2)}+{\cal O}(\as^3)$, and $a$ stands for either
$\as(\mu^2)$ or $\asbar$. The coupling $\asbar(k^2)$ is the
$d$-dimensional strong coupling constant
\cite{Magnea:1990zb,Magnea:2000ss}, known through NNLO
\cite{Moch:2005id}. It is a function of the usual four dimensional
coupling $\as(\mu^2)$ and the dimensional regulator $\ep$ (the
explicit relation we use here can be found in
Ref.~\cite{Mitov:2006xs}
\footnote{Note that in Refs.~\cite{Mitov:2006xs} and
\cite{Moch:2005id} the coupling $\asbar(k^2)$ is normalized to
$1/(4\pi)$ while in this paper we work in normalization
$\asbar/\pi$.}).
The result of Eq.\ (\ref{eq:S-function-amplitude}) depends on the
vanishing of the running coupling $\as(\mu^2)$ at $\mu=0$ for
$\ep<0$, that is, in more than four dimensions.

Note the similarity between Eq.~(\ref{eq:S-function-amplitude}) and
(\ref{eq:S-observ}): the amplitude term can be directly obtained
from one of the exponents in Eq.~(\ref{eq:S-observ}) by simply
ignoring the term with $x^{N-1}$ and replacing the four dimensional
coupling with the $d$-dimensional one. Such {\it duality} is not
accidental; physically both exponents can be thought of as two
different regularizations of the soft limit with regulators,
respectively, $\ep$ and $\ln(N)$. This relation has been explored
and detailed, for example, in Ref.~\cite{Mitov:2006xs}.

Without affecting the value of the soft anomalous dimension matrix,
one has the freedom to add finite $\ep$-terms in the soft function
which amounts to a re-definition of the hard function ${\bf h}$.
In Eq.\ (\ref{eq:S-function-amplitude})
 we choose a minimal, \MSbar -inspired scheme, where only $\ep$
poles are kept in the soft function. With this scheme defining the
separation between the soft and hard functions, the explicit relation
between the soft function and the anomalous dimension matrix through
two loops reads:
\begin{equation}
\label{eq:S-function-ampl-RESULT} \ln {\bf s}(\ep,\dots)  =
{\as(\mu^2)\over \pi} ~{{\bf\Gamma}_S^{(1)}\over 2\ep} +
\left({\as(\mu^2)\over \pi}\right)^2 \left( -{\beta_0
{\bf\Gamma}_S^{(1)}\over 16\ep^2} +{{\bf\Gamma}_S^{(2)}\over 4\ep} \right) \, .
\end{equation}

To summarize our discussion up to here, we have shown that the soft
function in Eq.~(\ref{eq:observable}) can be fully specified by
(\ref{eq:S-observ},\ref{eq:S-tilde}) in terms of the anomalous
dimension matrix ${\bf \Gamma}_S$ which in turn is derived solely
from the knowledge of the purely virtual corrections to the same
process. Ambiguities in ${\bf \Gamma}_S$ are fixed by choosing a
prescription at the level of the amplitude and we work with the form
factor prescription for both massless and massive partons. Once the
soft function in Eq.~(\ref{eq:observable}) has been fixed, in order
to perform NNLL resummation in observables, one has to determine the
number and the form of the various jet functions related to that
observable in a manner consistent with the prescription implicit in
the definition of the soft function. We turn to this next.

In the spirit of the form factor scheme that we employ here, we will
associate a jet function in (\ref{eq:amplitude}) for each of the
hard partons, {\it both massless and massive}. The number of hard
partons in the same underlying process determines also the number of
jet functions in the decomposition of an observable
(\ref{eq:observable}). To derive the expressions for various jet
functions in the following we use their process independence to
either  calculate them directly or to extract them from known
results.

\subsection{Jet functions for incoming partons}

We start with the jet function $J_{\rm in}$ for an initial state
hard parton (quark or gluon) which is basic for all hadron collider
processes. To this end, we can use the well known results from Drell-Yan
vector boson or Higgs boson production. In these reactions,
Eq.~(\ref{eq:observable}) takes the form
\begin{equation}
\label{eq:DY-Higgs} \sigma^P(N,Q) = \left[J_{\rm in}^P(N,Q)\right]^2
~ H(Q) S(N,Q) + {\cal O}(1/N) \, ,
\end{equation}
for $P\in\left(q \leftrightarrow {\rm DY},g \leftrightarrow {\rm
Higgs}\right)$. In the two processes, the hard scale $Q$ is simply
the virtuality of the outgoing color singlet
vector boson in DY or the mass
of the Higgs boson. Since in these two reactions exactly two hard
colored partons are involved, the hard and soft functions are just
$1\times 1$ matrices, i.e. the color structure is trivial. Upon
setting $\ln(\mu^2/s)=0$ in the results of Ref.~\cite{Aybat}, it
follows that the soft-anomalous dimension matrix vanishes through
two-loops (and possibly  to all orders
\cite{Aybat,Becher:2009cu,Gardi:2009qi,Becher:2009qa}). Therefore,
in Drell-Yan and Higgs boson production we have simply $S(N,Q)=1$.
Thus, the jet function $J_{\rm in}$ is simply the square root of the
corresponding Sudakov factors, see for example
Refs.~\cite{Moch:2005ba,Moch:2005ky}:
\begin{equation}
\label{eq:Jin} \ln J_{\rm in}^P(N,Q) = {1\over 2} \int_0^1
dx{x^{N-1}-1\over 1-x} ~ \Bigg\{ \int_{\mu_F^2}^{(1-x)^2Q^2}
{dq^2\over q^2}\, 2\, A_P\left(\as\left[q^2\right]\right) +
D_P\left(\as\left[(1-x)^2Q^2\right]\right) \Bigg\} \, .
\end{equation}
The functions $A_P, P=(q,g)$ and $D_P, P=(q,g)$ are currently known
through three loops (\cite{Moch:2004pa,Vogt:2004mw} and
\cite{Moch:2005ky,Laenen:2005uz,Becher:2007ty} respectively). The
factorization scale $\mu_F$ appearing in Eq.~(\ref{eq:Jin}) is
related to the factorization of the non-perturbative parton
distributions, assumed to be defined in the \MSbar\ scheme.
Utilizing a perturbative distribution function
\cite{Gardi:2005yi,Mele:1990cw,Melnikov:2004bm} one can also extend
that result to processes initiated by massive partons
\cite{Mitov:2006xs}.

The derivation of the jet functions for final state hard partons is
more involved since these depend on the definition of the
observable. Similarly to Drell-Yan/Higgs, one can use the vanishing
of the soft anomalous dimension matrix (and thus the absence of
non-trivial soft-gluon correlations) in any process involving two
hard colored partons in order to extract various jet factors. For
example, jet functions for ``observed" outgoing hard partons
(fragmentation) can be derived from semi-inclusive $e^+e^-$
annihilation to hadrons
\cite{Cacciari:2001cw,Rijken:1996ns,Mitov:2006wy,Sterman:2006hu}. Extension to the
massive case can be done in a fashion similar to the case of
Drell-Yan discussed above.

Of particular interest to us in this work are observables with
inclusive final states; a very well known example is inclusive DIS
\cite{dyresum,Moch:2005ba} which can be treated similarly to
Drell-Yan and $e^+e^-$, as discussed above. We are furthermore
interested in processes with non-trivial color correlations, like
the resummation of soft-gluons at NNLL in $t{\bar t}$
hadro-production. In order to calculate all jet factors that enter
that observable we need to first specify the soft anomalous
dimension matrix in this process which is done in section
\ref{sec:gamma2}. The calculation of the final state jet factors and
the final result for the cross section are relegated to section
\ref{sec:NNLL-res-crossection}.

Finally we would like to comment on the process independence of the
various jet factors discussed above. In principle, the presence of a
process dependent hard scale $Q$ indicates process dependence of the
whole result. What is process independent is the functional form of
the corresponding jet functions, while the dependence of the hard
scale should be thought of as a sort of functional argument related
to the phase-space for soft-gluon radiation available in the given
process. Therefore in different processes the ``argument" of the jet
functions will in general be different but their functional form
stays the same.

\section{From Differential to Inclusive Cross Sections}
\label{sec:reduction}

The resummed partonic hard scattering function $\omega_P(z)$   at fixed
invariant mass is found from its moments with respect to $z=M^2/s$,
with $M$ the pair invariant mass and $s$ the partonic center of mass
energy squared. The fully inclusive hard scattering cross section is
then found by integrating over  $M$, or equivalently, over $z$,
and the result is a function of
\bea
\rho \equiv \frac{4m^2}{s}
\eea
 only. We must also
integrate over the center-of-mass scattering angle (equivalently,
$\hat\eta$ above), but as we shall see, this does not affect our reasoning, and we
suppress this integral for simplicity of notation. In expressing our
results, we will find it useful to note that the ratio of pair and
particle masses obeys the relation \bea \frac{4m^2}{M^2} =
\frac{\rho}{z}\, . \label{rhoz} \eea

We denote the inclusive
cross section at parton level as $\sigma_P(\rho,s)$.  It is related
to the differential cross section in $z$
at fixed $s$ by simply integrating over $z$, with
lower limit $\rho$,
\bea
\sigma_P(\rho,s)
&=&
\int_\rho^1 dz\; \omega_P(z,M,m)\, .
\label{sigmarho}
\eea
We have observed that the singular dependence
of $\omega_P$ on $z$ can be found in turn by an
inverse transform,
\bea
\omega_P(z,M,m)
=
\int \frac{dN}{2\pi i} z^{-N}\ \omega_P^{\rm res}(N,M,m)\, .
\eea
The $z$-resummed cross section is taken at fixed
pair invariant mass $M$, and therefore fixed velocity
in the hard-scattering c.m.,
\bea
\beta_M^2 \equiv 1-\frac{4m^2}{M^2}\ =\ 1 - \frac{\rho}{z}\, .
\label{betadef}
\eea
Our goal is to relate the expression for $\omega^{\rm res}(N,M,m)$
at fixed $M$ to  the inclusive resummed cross section $\sigma_P(\rho,s)$ with
respect to $\rho$, as an inverse transform from moment
space in terms of that variable.

The resummed expression for $\omega(N)$ is given in
Eq.~(\ref{eq:observable}). For the following analysis, we make a
slight change in notation, and recognize that the hard-scattering
function ${\bf H}$, which describes the short-distance part of the
cross section, vanishes linearly in the center-of-mass velocity at
absolute threshold, and lowest order in $\as$.   Since $\beta_M$ depends only on the ratio $m/M$,
this quantity is fixed for Mellin moments with respect to $z$.   It
will, however, be important for $\sigma_P$. To make this trivial but
important factor explicit, we change the notation slightly, and
write \bea
{\bf H}^P\left(\frac{M^2}{\mu^2},\frac{m^2}{\mu^2},\as(\mu^2)\right)
\rightarrow
\beta_M\, {\bf H}^P\left(\frac{\rho}{z},\frac{M^2}{\mu^2},\as(\mu^2)\right)\, ,
\eea
so that
Eq.\ (\ref{eq:observable}) becomes
\bea
\omega_P^{\rm res}(N,M,m)
=
\beta_M\ \prod_i J_i(N)
\ {\rm Tr}\left[ {\bf H}^P\left(\frac{\rho}{z},\frac{M^2}{\mu^2},\as(\mu^2)\right)
{\bf S}^P\left(\frac{N^2\mu^2}{M^2},\frac{M^2}{m^2},\as(\mu^2)\right)\right]
\, ,
\eea
where we have represented the jet functions schematically.
We emphasize that moments in $z$ at fixed $M$ are equivalent
to integrals over $s$.   They thus leave $\beta_M$,
but not $\rho$, fixed.   Dependence on $\rho$ enters only after
the integral over $M^2$, or equivalently $z$, as in Eq.\ (\ref{sigmarho}).

We now relate the inclusive
cross section, which depends on $\rho$,
to the fixed-$M^2$ $z$ moments of $\omega$ by
\bea
 \sigma_P(\rho,s)
&=&
 \int_\rho^1 dz
\int \frac{dN}{2\pi i}\ z^{-N}\ \omega_P^{\rm res}(N,M^2,m^2)
\nonumber\\
&=&
\int \frac{dN}{2\pi i}\
\int_\rho^\infty dz\ z^{-N}\
 \sqrt{1-\frac{\rho}{z}}\,  \prod_i J_i(N)
\ {\rm Tr}\left[ {\bf
H}^P\left(\frac{\rho}{z},\frac{M^2}{\mu^2},\as(\mu^2)\right) {\bf
S}^P\left(\frac{N^2\mu^2}{M^2},\frac{z}{\rho},\as(\mu^2)\right)\right]
\,, \label{sigmarhoS} \eea where in the second form we have observed
that $\omega_P(z)$ vanishes for $z>1$, and have exchanged the $N$
and $z$ integrals. We note that when we choose $\mu=cm$, with $c$
some constant of order unity, the full right-hand side of Eq.\
(\ref{sigmarhoS}) depends on $M$ only through logarithms of the
ratio $4m^2/M^2=\rho/z$, see Eq.\ (\ref{rhoz}). Making such a choice
of renormalization scale and changing variable from $z$ to
\bea
\xi \equiv \frac{z}{\rho}\ =\ \frac{1}{1-\beta_M^2}\, ,
\label{xidef}
\eea
we derive the desired form of an inverse
transfrom, \bea
 \sigma_P(\rho,s)
&=&
\int \frac{dN}{2\pi i} \rho^{-N+1}\ \sigma_P(N,m)\, ,
\eea
with
\bea
\sigma_P(N,m)
&\equiv&
\int_1^\infty d\xi\ \xi^{-N}\
 \sqrt{1-\xi^{-1}}\, \prod_i J_i(N,\xi,\as)
\ {\rm Tr}\left[ {\bf H}^P\left(\xi,\as(\mu^2)\right)
{\bf S}^P\left(N^2\xi,\xi,\as\right)\right] \, .
\label{xiintegral}
  \eea
  Here we have simplified the notation for the arguments of
  the jet, hard and soft functions somewhat to emphasize their $\xi$-dependence.
  The scale of the running coupling is, as indicated above,
  of order of the quark mass, $m$.
The relationship between the $N$-dependence of the resummed cross
section at fixed $M$ in $\omega_P^{\rm res}(N)$ and in
$\sigma_P(N,m)$ can be found readily in the large-$N$ limit, by
noting that the integral over $\xi$ in (\ref{xiintegral}) is
dominated by the
factor
\bea \xi^{-N} \sim  e^{N\ln(1-\beta_M^2)}\, ,
\eea
which forces $M^2$ towards $4m^2$. Center-of-mass velocities
$\beta_M \gg 1/\sqrt{N}$ are thus exponentially suppressed.
Correspondingly, the scale of energy evolution in the soft cross
section, Eq.\ (\ref{eq:S-observ}) is over an interval from $m$ to
$m/N\ge m\beta_M^2$.   For this range of energies, the
evolution variable $\mu'$ in (\ref{eq:S-observ}) is larger than the
kinetic energy of the pair in their center of mass. We shall assume
below that, as suggested in Ref.\ \cite{Beneke:2009rj}, radiation in
this energy range decouples from the pair, whose interactions give
rise to Coulomb enhancements that appear as inverse powers of
$\beta_M$. In the soft anomalous dimension matrix appropriate to
this range of energies, the pair of heavy quark eikonals is
effectively replaced by a singlet or octet eikonal line, with a
separate term that describes the evolution of the pair. This
approximation results in a smooth limit at absolute threshold
$\beta_M=0$. \footnote{We will discuss the relationship of this
framework to the very interesting and recent results of Ref.\
\cite{Ferroglia:2009ep} elsewhere. \cite{CMSI-prep}]}

Corrections due to the logarithmic
$\xi$-dependence in the jet and soft factors are suppressed by
inverse powers of $N$.
Up to such corrections, the result is the
Born cross section for heavy quark production in process $P$ times
the remaining jet, hard and soft functions, which we write as
\bea
\sigma_P(N,m) = \sigma_{\rm Born}^P(N)\ \prod_i J_i(N,1,\as) \ {\rm
Tr}\left[ {\bf \hat{H}}^P\left(1,\as(\mu^2)\right) {\bf
S}^P\left(N,1,\as\right)\right] \left ( 1 +{\cal O}\left(\frac{1}{N}\right)\right)\, ,
\label{sigmaPresult}
\eea
where the hat on the hard matrix indicates
that have factored out the $N$-dependence of the Born cross section,
which behaves at leading power in $N$ as
\bea
\sigma^P_{\rm Born}(N) \sim
\int_1^\infty d\xi\, \xi^{-N+1/2}\, \sqrt{\xi-1} =
\frac{\sqrt{\pi}}{2}\, \frac{1}{N^{3/2}}\, \left ( 1 +{\cal O}\left(\frac{1}{N}\right)\right)\, .
\eea
Eq.\ (\ref{sigmaPresult})
is the form that we will use below.
As suggested above,
we will evaluate the soft function ${\bf S}^P(N,1,\as)$ using
Eq.\ (\ref{eq:S-observ}) computed with a soft anomalous dimension
appropriate to the range energy range $m>\mu' > \beta_M^2m$.

\section{The two-loop anomalous dimension matrix at absolute threshold}
\label{sec:gamma2}

In this section we derive the relevant result for the two-loop
anomalous dimension matrix ${\bf\Gamma}_S$.
We also show that these results are enough to predict the
full pole structure of the two-loop color averaged amplitudes
for $q\bar{q}\rightarrow t\bar{t}$.

The one-loop massive anomalous dimension matrix for an amplitude
with $n$ colored partons, ${\cal N}_m$ of which are massive and with
equal mass $m$ has been known for some time
\cite{Kidonakis:1997gm,Catani:2000ef}:
\begin{equation}
{\bf \Gamma}_S^{(1)} = {1\over 2} \sum_{(i\neq j) = 1}^n T_i\cdot T_j
\ln\left( -{\mu^2\over \s_{ij}}\right) + {1\over 2}\sum_{(i\neq j)
\in {\cal N}_m} ~ T_i\cdot T_j ~ \left[ \ln\left(1+\xij^2\right) +
{2\xij^2\over 1-\xij^2}\ln(\xij) \right] \, ,
\label{eq:Gamma1mass-result}
\end{equation}
where $s_{ij} = (p_i+p_j)^2$ and $\s_{ij}=2p_i\cdot
p_j=s_{ij}-m_i^2-m_j^2$ (with $m_{i,j}=\{0,m\}$). The space-like
variables $\xij$ read \cite{Bernreuther:2004ih}:
\begin{equation}
\label{eq:x-y-deff} {m^2\over s_{ij}} = - {\xij\over (1-\xij)^2} ~~~
, ~~~ \xij = { \sqrt{1-{4m^2\over s_{ij}}} - 1 \over
\sqrt{1-{4m^2\over s_{ij}}} + 1} \,\, ,
\end{equation}
when, in the unphysical space-like kinematics, all invariants
$s_{ij} <0$. In specific applications some of them have to be
continued to time-like kinematics. This can be done with the help of
the replacement $\xij = - \yij + i\ep$, where $s_{ij}$ is now in the
physical region $ s_{ij}\geq 4m^2$ and the ``time-like'' variable is
$0<y_{ij}\leq 1$. The color generators $T_i$ are defined such that
they satisfy $\sum_k^n T_k = 0$, and can be either in the
fundamental or adjoint representation of the color group for quarks
or gluons. The index $i$ labels the leg where the generator is
inserted; see also appendix \ref{sec:app} for more details.

In parallel to the two-loop massless case \cite{Aybat}, the two-loop
massive anomalous dimension matrix ${\bf \Gamma}_S^{(2)}$ is built up
from 2- and 3-eikonal (3E) contributions, i.e. configurations where soft
gluons are exchanged between two (resp. three) external hard
partons. 
\begin{equation}
{\bf \Gamma}_S^{(2)} = {1\over 2} \sum_{(i\neq j) = 1}^n T_i\cdot
T_j~{K\over 2}~\ln\left( -{\mu^2\over \s_{ij}}\right) + {1\over
2}\sum_{(i\neq j) \in {\cal N}_m} ~ T_i\cdot T_j ~ P^{(2)}_{ij} +
3E\; {\rm terms} \, , \label{eq:Gamma2mass-ansatz}
\end{equation}
where, as for the massless case, $K = ( 67/18 - \pi^2/6)C_A -
(5/9)N_L$.
Even before treating the 3E terms, we see that
at two loops exchanges involving two eikonals
take on the same color structure as in the one-loop
anomalous dimension, Eq.\ (\ref{eq:Gamma1mass-result}).

Even without using explicit forms for the $3E$ contributions \cite{Ferroglia:2009ep}, we have adequate information
to study the behavior of the soft anomalous dimension
in the range $\mu>\beta^2m$, subject to our assumption
of factorization, as discussed above.   In \cite{Mitov:2009sv}
it was observed, for example, that 3E contributions
to the reactions $q{\bar q} \to Q{\overline Q}$ and $gg \to Q{\overline Q}$ vanish,
either identically (when two eikonal lines are massless) or at $u=t$
(for two massive eikonals).
Given our assumption of the decoupling of soft radiation
from the dynamics of the pair in the range $\mu'> \beta^2m$,
we may extend the anomalous dimension,
appropriate for this range of energies to absolute threshold,
 $\beta=\sqrt{1-4m^2/s}\to 0$.   To the order at which we
 work, power singularities in $\beta$ associated with
 independent evolution of the pair of heavy eikonal lines cancel
 in the soft function $S$ of Eq.\ (\ref{eq:S-observ}).
It is this simplification that enables us to present a
full expression for the threshold-resummed inclusive cross
section at NNLL.  To derive this result, we need only the 2E diagrams.

The $2E$, dipole-type contributions, can be readily derived in
complete generality.
Here we note first that Eq.~(\ref{eq:Gamma2mass-ansatz}) reproduces
the known massless result for a massless dipole. Second, it reflects
the fact that, similarly to the one loop case, the mixed corrections
between massive and massless legs do not produce any power
corrections in $m^2/s_{ij}$. To verify that one-mass dipoles do not
involve additional power corrections, we have repeated for this case
the arguments given below for the derivation of the function
$P^{(2)}_{ij}$. In that check we have used the recent two-loop
calculation of the heavy-to-light form factor in QCD
\cite{Bonciani:2008wf,Beneke:2008ei}; see also
\cite{Bell:2006tz,Asatrian:2008uk,Huber:2009se}. For partial checks
we have made use of the packages HPL \cite{HPL} and FIESTA
\cite{Smirnov:2008py}. Note also that the absence of power
corrections in the mass in the one-mass dipoles is related to the
choice of the variable $\s_{ij}=2p_i\cdot p_j$ instead of
$s_{ij}=(p_i+p_j)^2$ in the first term of
Eqs.~(\ref{eq:Gamma1mass-result},\ref{eq:Gamma2mass-ansatz}).

Finally, we explain how the functions $P^{(2)}_{ij}$ can be
determined. The dependence on the indices $(i,j)$ of the function
$P^{(2)}_{ij}$ in Eq.~(\ref{eq:Gamma2mass-ansatz}) is only through
the corresponding kinematical invariant $s_{ij}$, i.e. $P^{(2)}_{ij}
= P^{(2)}(s_{ij})$, and the dependence on $s_{ij}$ enters through
the variable $\xij$ defined in Eq.~(\ref{eq:x-y-deff}). That implies
its functional form is universal and therefore can be extracted from
the simplest two-loop amplitude with $n=2$: the two-loop massive
vector form factor $F_1\left( \gamma^*\to Q {\overline Q}\right)$
\cite{Bernreuther:2004ih,Gluza:2009yy}. In this case
Eq.~(\ref{eq:Gamma2mass-ansatz}) simplifies to
\begin{equation}
{\bf \Gamma}_S^{(2)}(n=2) = - C_F \left[ {K\over 2}~\ln\left( -{\mu^2\over
\s}\right) + P^{(2)}(s) \right]\, , \label{eq:Gamma2mass-FF}
\end{equation}
where $\s = s-2m^2$ and $s=(p_1+p_2)^2 < 0$. Of course, there are
no $3E$ contributions for $n=2$.

As we remarked above, the soft anomalous dimension matrix is
defined only up to a term proportional to the unit matrix.
In this context, we can use that ambiguity in the
definition of the soft function to define it through the condition
$H=1$ in the factorization of the form factor $F_1 = J\cdot S \cdot
H$ following from Eq.~(\ref{eq:amplitude}). From the known results
for the form factor $F_1$ and the jet function $J$
\cite{Mitov:2006xs} and taking into account
Eq.~(\ref{eq:S-function-ampl-RESULT}) we derive:
\begin{equation}
P^{(2)} = {K\over 2} P^{(1)} + P^{(2),{\rm m}}\, ,\label{eq:P2mass}
\end{equation}
where, similarly to the definition of $P^{(2)}$ in
Eq.~(\ref{eq:Gamma2mass-ansatz}), the function $P^{(1)}$ equals the
term in the square brackets in Eq.~(\ref{eq:Gamma1mass-result}). The
presence of the term $P^{(2),{\rm m}}$ in the above equation
indicates that the property of the two loop massless amplitudes
$\Gamma^{(2)} = K/2~ \Gamma^{(1)}$ \cite{Aybat} is broken in the
massive case by power corrections of the mass. The function
$P^{(2),{\rm m}}$ is given by
\begin{eqnarray}
P^{(2),{\rm m}}(x) &=& {C_A\over (1-x^2)^2}~\Bigg\{ -{(1+x^2)^2\over
2} {\rm Li}_3(x^2) + \left( {(1+x^2)^2\over 2} \ln(x) -{1-x^4\over
2}\right)  {\rm Li}_2(x^2) \nonumber\\
&+& {x^2(1+x^2)\over 3} \ln^3(x) + x^2(1-x^2) \ln^2(x)\nonumber\\
&+& \left( - (1-x^4) \ln\left(1-x^2\right) + x^2(1+x^2) \zeta_2
\right) \ln(x)  + x^2(1-x^2) \zeta_2 + 2x^2\zeta_3 \Bigg\} \, ,
\label{eq:P2new}
\end{eqnarray}
where $\zeta_n$ is the Riemann zeta function: $\zeta_2 = \pi^2/6,~
\zeta_3 = 1.202057\dots$.

The function $P^{(2),{\rm m}}$ (and in particular its real part)
does not vanish at threshold; for example, for a time-like argument
$x=-(1-\beta)/(1+\beta) + i\varepsilon$ with $\beta=\sqrt{1-4m^2/s}$
and $0\leq \beta \leq 1$, this limit is
\begin{equation}
P^{(2),{\rm m}}(x\sim -1+i\varepsilon) = {1-\zeta_3\over 2}C_A\ +
\left( {\pi^2\over 24}-{1\over 2}\right){i\pi\over \beta} C_A +
{\cal O}(\beta)\, .
\label{eq:P2new-threshold}
\end{equation}
This result contains, as usual, a Coulomb enhancement in
its imaginary part, which reflects the pair's internal evolution.

Combining the results above, the two-loop soft anomalous dimension
matrix for the two-to-two quark- and gluon-initiated reactions (see
Eqs. (\ref{eq:qq-QQ}) and (\ref{eq:gg-QQ}))
takes the following form close to absolute threshold $\beta\to 0$,
\begin{equation}
{\bf \Gamma}_S^{(2)} = {K\over 2}~{\bf \Gamma}_S^{(1)} + T_3\cdot
T_4 ~ P^{(2),{\rm m}}_{34} ~~,~~ ({\rm for}~\beta\to 0)\, ,
\label{eq:Gamma2mass-qq-gg}
\end{equation}
with $P^{(2),{\rm m}}_{34}$ given by Eq.~(\ref{eq:P2new-threshold}).
The explicit results for the matrices ${\bf \Gamma}_S^{(1)}$ and
$T_3\cdot T_4$ can be found in appendix \ref{sec:app}.

The mass-dependent soft anomalous dimension of
Eq.\ (\ref{eq:Gamma2mass-FF}) for processes with
the color structure of the form factor
was derived  in Ref.\ \cite{Kidonakis:2009ev},
using a slightly different scheme for the soft function.
Eq.~(\ref{eq:Gamma2mass-ansatz})
for soft matrices of arbitrary $n$-point amplitudes involving massive colored particles
was presented in Ref.~\cite{Becher:2009kw}.
To determine the analogue of the function
$P^{(2),{\rm m}}$, the authors of that reference have utilized the
results of Refs.~\cite{Korchemsky:1987wg} and \cite{Kidonakis:2009ev}. We find
agreement between our $P^{(2),{\rm m}}$ and ($-1$ times) the function appearing
in Eq.~(15) of version 2 of the arXiv preprint of
Ref.~\cite{Becher:2009kw}.
The fact that the results of
Ref.\ \cite{Becher:2009kw} reproduce the IR poles of the massive form
factor (which we have used to extract the function $P^{(2),{\rm
m}}$) implies a non-trivial consistency between our results and the
results of Refs.~\cite{Kidonakis:2009ev,Becher:2009kw}.

Moreover we have performed for the first time a truly non-trivial
check of Eq.~(\ref{eq:Gamma2mass-ansatz}) as a whole, by predicting
the IR poles of the squared two-loop $q{\bar q} \to Q{\overline Q}$
amplitude and comparing them to the numerical calculation of
Ref.~\cite{Czakon:2008zk}. We have found a complete agreement
between the predictions of our formalism and the color-averaged
squared amplitude at two loops.

In order to be able to make this
prediction, we have noticed that in squaring the amplitude and
summing/averaging over colors, any $3E$-type contributions in
Eq.~(\ref{eq:Gamma2mass-ansatz}) with color structure $f^{abc} T^a
T^b T^c$ would vanish simply by color projection, and would not
contribute to the squared amplitude at this level.    Thus the
calculation in question does not test for the presence of
such terms at the amplitude level.

The setup of our prediction is as follows: the amplitude $M$,
multiplied by the Born amplitude and summed over spin/color, can be
expanded in the coupling $a_{\rm s} = \as(\mu^2)/(2\pi)$ as
\begin{eqnarray}
M = M^{(0)}(\ep) + a_{\rm s} M^{(1)}(\ep) + a_{\rm s}^2 M^{(2)}(\ep)
+ {\cal O}\left(a_{\rm s}^3\right) \, . \label{eq:JSH}
\end{eqnarray}
The factorization properties of on-shell amplitudes detailed in
section \ref{schemesec} give the following prediction for the poles
of the amplitude $M$ through two loops
\begin{eqnarray}
M^{(1)}(\ep) &=& \Bigg\{{1\over \ep}\Gamma_1+J^{(1)}\Bigg\}~ M^{(0)}
+ {\cal O}(\ep^0) \, ,\nonumber\\
M^{(2)}(\ep) &=& \Bigg\{J^{(2)} -\left(J^{(1)}\right)^2+ {1\over
\ep}\left(-J^{(1)} \Gamma_1+\Gamma_2\right) + {1\over
\ep^2}\left(-{1\over 2}(\Gamma_1)^2-{\beta_0\over
4}\Gamma_1\right)\Bigg\}~M^{(0)}
\nonumber\\
&+& \Bigg\{{1\over \ep}\Gamma_1+J^{(1)}\Bigg\}~ M^{(1)} + {\cal
O}(\ep^0) \, . \label{M2poles}
\end{eqnarray}
The function $J(\ep)$ represents the product of the four $\ep$-dependent jet
functions corresponding to the two incoming massless and two
outgoing massive fermions (see Refs.~\cite{Czakon:2007ej} for more
details) and has an expansion in $a_{\rm s}$ analogous to
Eq.~(\ref{eq:JSH}). Similarly, $\Gamma_1$ and $\Gamma_2$ are the
expansion in terms of the coupling $a_{\rm s}$ of the anomalous
dimension matrix ${\bf\Gamma}_S$ given through
Eqs.~(\ref{eq:Gamma1mass-result},\ref{eq:Gamma2mass-ansatz}).

Predicting all two-loop poles in the squared amplitude requires also
the one-loop amplitude for the same process evaluated to
sufficiently high order in $\ep$. We have calculated them
separately. Our results will provide a non-trivial check on future
extension of the results of Refs.~\cite{Bonciani:2008az} to the
analytic calculation of the non-planar contributions in this
process.

\section{NNLL Resummation for total $t{\bar t}$ hadro-production cross section}
\label{sec:NNLL-res-crossection}

We can summarize the results of sections \ref{sec:soft-gluon-res}
and \ref{sec:reduction} for the resummed
partonic total-inclusive cross section for
$t{\bar t}$ pair production at hadron colliders in moment space by
\begin{equation}
\label{eq:sigma-top} \sigma^P(N,m^2,\mu^2)
=
\sigma_{\rm
Born}^P(N)\; \left[J_{\rm in}^P(N,m^2,\mu^2)\right]^2~\left[J_{\rm
incl}(N,m^2,\mu^2)\right]^2 ~{\rm Tr}\left[ {\bf \hat{H}}^P(m^2,\mu^2)
{\bf S}^P(N,m^2,\mu^2)\right] + {\cal O}(1/N) \, .
\end{equation}
The index $P=(q,g)$ labels the jet functions as well as the
two reactions $q{\bar q} \to
t{\bar t}$ and $gg\to t{\bar t}$. The factorization/renormalization
scales are denoted by $\mu$. The jet factors $J_{\rm in}^P$ are
given in Eq.~(\ref{eq:Jin}) with $Q^2=4m^2$, where $m$ is the pole
mass of the top quark.
The only factors remaining to compute in (\ref{eq:sigma-top}) are functions $J_{\rm incl}$
for the final-state jets
and the one-loop correction to the soft function at $N=1$ in
Eq.\ (\ref{eq:S-observ}), which serves as a boundary
condition for the evolved soft function.   We turn first to the outgoing jets.

The outgoing jet functions are specified by our choice of the form factor scheme,
as described in Sec.\ \ref{schemesec}.    Their virtual contributions
precisely cancel the terms subtracted from the soft anomalous dimension matrix ${\bf \Gamma}_S$
in Eq.\ (\ref{eq:Gamma1mass-result}), and hence in the soft function ${\bf S}$ appearing in
Eq.~(\ref{eq:sigma-top}).   Specifically, we have subtracted those
soft  singularities corresponding to the low-mass limit
of the outgoing legs.
For a completely inclusive observable, like the total inclusive
cross section, such factorization is not strictly necessary.
The resulting expression, however, Eq.\ (\ref{eq:sigma-top}), provides
a unified threshold resummation for the total inclusive
cross section and for the cross section at measured pair
invariant mass $s\ge 4m^2$, including the limit $s\gg m^2$,
where logarithms of the heavy quark mass can be important.

As described above, the outgoing jet function can be constructed
directly from the exponentiation of its infrared singularities in
the low-mass limit, and therefore is of the form,
\begin{equation}
\label{eq:J-incluisve} J_{\rm incl}(N,m^2,\mu^2) =
\exp\Bigg\{{1\over 2}\int_0^1 dx{x^{N-1}-1\over 1-x} ~ \Gamma_{\rm
incl}\left(\as\left[4m^2(1-x)^2\right]\right) \Bigg\} \, .
\end{equation}
The jet anomalous dimension $\Gamma_{\rm incl}$ is
proportional to the unit matrix in color space.
Specifically, it is given by the single
poles of the logarithm of the massive quark form factor in the small mass limit,
which defines the jet factor of a massive line in an amplitude
\cite{Mitov:2006xs}, and which we have adapted here for the
form factor scheme.
The explicit expression for $\Gamma_{\rm incl}$
through two loops is
\begin{equation}
\label{eq:Gamma-inclusive} \Gamma_{\rm incl}  = {\as(\mu^2)\over
\pi} ~C_F \left[-1-\ln\left({m^2\over \mu^2}\right)\right] +
\left({\as(\mu^2)\over \pi}\right)^2 \left[ {K\over 2}C_F\,
\left(-1-\ln\left({m^2\over \mu^2}\right)\right) - {\zeta_3-1 \over
2} C_FC_A \right] \, .
\end{equation}
The non-logarithmic part of $\Gamma_{\rm incl}$ can be naturally
expressed in terms of the anomalous dimensions $G,K$ needed for the
exponentiation of the massive form factor to NNLL
\cite{Mitov:2006xs},
\begin{equation}
\label{eq:Gamma-inclusive-G-and-K} \Gamma_{\rm incl}({\rm non-log\
term}) = {\as(\mu^2)\over \pi} ~{1\over 4}~\left[G_1^{(0)} +
K_1\right] + \left({\as(\mu^2)\over \pi}\right)^2~{1\over
4^2}~\left[G_2^{(0)} + K_2 - \beta_0G_1^{(1)}\right] \, .
\end{equation}
The functions $G$ and $K$ are defined in \cite{Mitov:2006xs} as
expansions of $\as/(4\pi)$, hence the additional powers of $1/4$ in
the equation above, and with $N_L$ active flavors. The function $G$
is $\ep$-dependent: $G_n = \sum_{i=0} G_n^{(i)} \ep^i$, and it
equals (minus) the function $G$ in the massless form factor
\cite{Magnea:1990zb,Magnea:2000ss,Moch:2005id}. The origin of the
term $\beta_0G_1^{(1)}$ can be understood along the lines of
Ref.~\cite{Dixon:2008gr}.

The last step remaining is the derivation of the boundary condition
${\bf S}^{(1)}\left(1,\beta_i\cdot\beta_j\right)$ for the soft
function ${\bf S}$, see section \ref{sec:soft-matrix}. We recall
that the boundary condition is uniquely defined once the form factor
scheme has been adopted. To extract it, we need to calculate the
total inclusive cross section in the eikonal approximation. After
the appropriate eikonal jet functions have been factored out (see
Ref.~\cite{Kidonakis:1997gm}) we are left with the desired boundary
condition.

The required one loop calculation is in fact quite straightforward.
To that end one can use the factorization in the soft limit of the
squared one-gluon real emission amplitude into the square ${\cal
S}_{ij}$ of the eikonal current and the Born amplitude with
appropriate insertions of the color operators $T_i\cdot T_j$ summed
over all pairs of legs $(i,j)$; see, for example,
Ref.~\cite{Catani:1999ss} for details. Combining matrix element
factorization with the factorization of phase space in the soft
limit, we arrive at $\sigma^{(1),real}_{eikonal}=\sum_{i,j=1}^4~
{\rm Born}^{ij} \times I_{ij}$. We label the legs according to the
momenta of the hard partons; see Eq.~(\ref{eq:gg-QQ}). The functions
$I_{ij}$ are simply the integrals of the eikonal current squared
over the phase space of the emitted gluon. While the integrand is
scaleless by construction, the integrals do not vanish because we
integrate up to the maximal energy $E_{max}$ available to the
emitted gluon in the partonic c.m.\ system. Their expressions read:
\begin{equation}
\label{eq:J-ij} I_{ij} = {\as(\mu^2)\over \pi}\left( {\mu^2\over
4E_{max}^2}\right)^\ep ~ J_{ij}\, , ~~~ {\rm where:}~~~ J_{ij} =
-{e^{\ep\gamma_E} \over 2^{2-2\ep}\pi^{1-e}}~\int_0^1 E_g^{1-2\ep}dE_g
\int d\Omega_{d-1} {(p_i\cdot p_j) \over (p_i\cdot g)(p_j\cdot g)}
\, .
\end{equation}

Working out the color algebra we get the following result for the
one-loop real-emission contribution for the reaction $gg\to Q{\bar
Q}$ (which covers the general case),
\begin{eqnarray}
\label{eq:sigma-1-eikonal} \sigma^{(1), real}_{eikonal} &=& {\bf
\sigma}_{\rm Born} ~ {\as(\mu^2)\over \pi}\left( {\mu^2\over
4E_{max}^2}\right)^\ep \nonumber\\
&\times&
\left[-2\left( C_F\left( J_{34}-J_{33}\right) + C_A J_{12}\right)
\left(
  \begin{array}{ccc}
   1  & 0 & 0\\
   0  & 1 & 0\\
   0  & 0 & 1\\
  \end{array}
\right)
+ C_A\left(J_{12} + J_{34} -2J_{13}\right)
\left(
  \begin{array}{ccc}
   0  & 0 & 0\\
   0  & 1 & 0\\
   0  & 0 & 1\\
  \end{array}
\right) \right] \, ,
\end{eqnarray}
where ${\bf \sigma}^{\rm Born}$ above is a diagonal matrix; we
work in the singlet-octet basis given in the appendix \ref{sec:app}.

Eq.~(\ref{eq:sigma-1-eikonal}) is derived in the back-to-back
scattering configuration where $u=t$ and holds for any
$\beta=\sqrt{1-4m^2/s}$. Nicely, the result is diagonal and the two
octets are degenerate. To complete this result one has to add the
corresponding virtual corrections. Since they are all scaleless and
thus vanish in dimensional regularization, the only contributions
comes from their counterterms.

The result simplifies significantly if one takes the limit $\beta\to
0$ which is relevant for the resummation of the total cross section.
In that limit (i.e. working up to powers of $\beta$) we can set
$\beta=0$ everywhere in the integrals $J_{ij}$. In this limit we
have $J_{13}=J_{12}/2$ and $J_{33}=J_{34}$ as well as
$4E_{max}^2=s\beta^4$. The two independent integrals read:
\begin{eqnarray}
\label{eq:J-12} J_{12} &=&  {e^{\ep\gamma_E}\over \Gamma(1-\ep)}
{1\over 2\ep} {\Gamma(-\ep)\Gamma(1-\ep)\over \Gamma(1-2\ep)} =
-{1\over
2\ep^2}+{\pi^2\over 8} + {\cal O}(\ep) \, , \\
\label{eq:J-34}
J_{34} &=& {e^{\ep\gamma_E}\over \Gamma(1-\ep)}
{1\over 4\ep} {2^{2\ep}\sqrt{\pi}\Gamma(1-\ep)\over \Gamma(3/2-\ep)}
= {1\over 2\ep} + 1 + {\cal O}(\ep) \, .
\end{eqnarray}
After subtracting the eikonal jets in such a way that the singlet
eigenvalue vanishes (i.e., to reproduce the well known result from
Drell-Yan-type processes) we finally get,
\begin{eqnarray}
\label{eq:sigma-1-eikonal-beta=0} \sigma^{(1)}_{eikonal} &=& {\bf
\sigma}_{\rm Born} ~ {\as(\mu^2)\over \pi}~C_A\left[ 1+ {1\over
2}\ln\left( {\mu^2\over 4E_{max}^2}\right) \right] \left(
 \begin{array}{ccc}
  0  & 0 & 0\\
  0  & 1 & 0\\
  0  & 0 & 1\\
 \end{array}
\right)\, .
\end{eqnarray}
The constant coefficient above follows from the constant term in
Eq.~(\ref{eq:J-34}). We have verified that
Eq.~(\ref{eq:sigma-1-eikonal-beta=0}) correctly reproduces the
$\ln\beta$ terms in the color-singlet color-octet difference of the
total inclusive cross-section (see Eqs.~(8-10) in
Ref.~\cite{Czakon:2008cx}).

Following the discussion in section \ref{sec:reduction} we set the
ratio $\mu^2/4E_{max}^2$ to unity, so that the logarithmic term in
Eq.~(\ref{eq:sigma-1-eikonal-beta=0}) vanishes. This way we get the
following result for the boundary condition of the soft function in
Eq.~(\ref{eq:S-observ}) relevant for the resummation of the total
heavy-pair cross-section,
\begin{eqnarray}
\label{eq:S-tilde_oneloop} {\bf S}
\left(1,\as\left(Q^2/N^2\right)\right) &=& {\bf S}^{(0)} \ \left[ 1
\ + \frac{\as\left(Q^2/N^2\right)}{\pi}\ C_A \left(
  \begin{array}{ccc}
   0  & 0 & 0\\
   0  & 1 & 0\\
   0  & 0 & 1\\
  \end{array}
\right)
+ \dots \right]
\nonumber\\
&=& {\bf S}^{(0)} \ \left[ 1 \  + \
C_A\frac{\as\left(\mu^2\right)}{\pi}\left\{ 1\   +
\frac{\as\left(\mu^2\right)}{\pi}\, {\beta_0\over 4} \ln
\left(\frac{N^2\mu^2}{Q^2}\right) \right\}
\left(
  \begin{array}{ccc}
   0  & 0 & 0\\
   0  & 1 & 0\\
   0  & 0 & 1\\
  \end{array}
\right)
+ \dots\right]\, ,
\end{eqnarray}
which as shown results in a term $\beta_0\, (C_A/2)\, \ln N$ in the
NNLL result. The ${\cal O}(\as)$ correction appears only when the
pair is produced in an octet configuration at short distances. The
result for $q{\bar q}\to Q{\overline Q}$ follows from
Eq.\ (\ref{eq:replacement}).

We are now ready to combine our previous findings and present our
result for the resummed heavy quark cross section in moment space up
to NNLL. Working in the singlet/octet basis for the soft anomalous
dimension ${\bf\Gamma}_S$, where it is diagonal \cite{Mitov:2009sv},
the result for the resummed cross section for heavy pair
hadroproduction reads
\begin{eqnarray}
\label{eq:sigma-top-TOT}
\frac{\sigma^P(N,m^2,\mu^2)}{\sigma_{\rm
Born}^P(N)} &=& {\rm Tr}\left[ {\bf \hat{H}}^P(m^2,\mu^2) {\bf
S}_P^{(0)}~\left[ 1~ + ~ \frac{\as\left(Q^2/N^2\right)}{\pi}\
C_A~{\bf \Pi}_8 \right]\ \exp\Bigg\{ \int_0^1 dx{x^{N-1}-1\over 1-x}
\right. \nonumber\\
&\times& \left.\left( \int_{\mu_F^2}^{4m^2(1-x)^2} {dq^2\over q^2}\,
2\, A_P\left(\as\left[q^2\right]\right)\,{\bf 1} +
{\hat D}^P_{Q{\overline Q}} \left(\as\left[4m^2(1-x)^2\right]\right)
\right)\Bigg\}\right] + {\cal O}(1/N,N^3LL)\, ,
\eea
where ${\bf \Pi}_8$ projects onto the color octet states (see Eq.\ (\ref{eq:replacement})),
and where ${\hat D}^P_{Q{\overline Q}}$ contains single-logarithmic
anomalous dimensions from both the jet and soft functions, in a color-diagonal
form (see below).
This expression is our central result.   It may be cast in a more familiar form,
by combining the constant piece of the soft function into the
hard function, and generating the $\ln N$ dependence in the
soft function from a slightly modified version of the function ${\hat D}^P_{Q{\overline Q}}$,
which we denote as simply $D^P_{Q{\overline Q}}$, and
which includes a new term proportional to $\as^2\beta_0$,
\bea
\frac{\sigma^P(N,m^2,\mu^2)}{\sigma_{\rm
Born}^P(N)}
&=& {\rm Tr}\left[ {\bf  H}^P(m^2,\mu^2)\ \exp\Bigg\{
\int_0^1 dx{x^{N-1}-1\over 1-x}
\right. \nonumber\\
&\times& \left.\left( \int_{\mu_F^2}^{4m^2(1-x)^2} {dq^2\over q^2}\,
2\, A_P\left(\as\left[q^2\right]\right)\,{\bf 1} +
D^P_{Q{\overline Q}} \left(\as\left[4m^2(1-x)^2\right]\right)
\right)\Bigg\}\right] + {\cal O}(1/N,N^3LL) \, ,
\label{eq:sigma-top-TOT-standard}
\end{eqnarray}
where
\bea
D^P_{Q{\overline Q}} = D_P\,{\bf 1} + 2
{\rm Re}\,{\bf\Gamma}_P + 2 \Gamma_{\rm incl}\,{\bf 1}
- \frac{1}{2}\, \left(\frac{\as}{\pi}\right)^2\, C_A\beta_0 \left(
  \begin{array}{cc}
   0  & 0 \\
   0  & 1 \\
  \end{array}
\right)\ .
\eea
The explicit form for $D^P_{Q{\overline Q}}$
uses the results for ${\bf \Gamma}_P$ in section \ref{sec:gamma2} and
appendix \ref{sec:app}.    In the reactions $q{\bar q}\to Q{\overline
Q}$ and $gg\to Q{\overline Q}$, which we label respectively by
$P\in\left(q\bar q,\, gg\right)$, it reads through two loops:
\begin{eqnarray}
D^P_{Q{\overline Q}} &=&
{\as(\mu^2)\over \pi} (-C_A)
\left(
  \begin{array}{cc}
   0  & 0 \\
   0  & 1 \\
  \end{array}
\right)
+ \left({\as(\mu^2)\over \pi}\right)^2
\Bigg\{
D^{(2)}_{\rm P}
\left(
  \begin{array}{cc}
   1  & 0 \\
   0  & 1 \\
  \end{array}
\right)
+\left(-C_A{K\over 2}-{\zeta_3-1\over 2}C_A^2 - C_A{\beta_0\over 2}\right)
\left(
  \begin{array}{cc}
   0  & 0 \\
   0  & 1 \\
  \end{array}
\right) \Bigg\} \, ,
\label{eq:D-QQ}
\nonumber\\
\end{eqnarray}
where $D_{q\bar q}=D_{\rm DY}$ and $D_{gg}=D_{\rm Higgs}$ \cite{Moch:2005ba}.
Corrections to Eq.\ (\ref{eq:sigma-top-TOT-standard}) begin, as indicated, at next-to-next-to-next-to
leading logarithm, and are determined by the three loop soft anomalous
dimension matrix and the inclusive soft function at two loops.

The hard function ${\bf  H}(m^2,\mu^2)$ in
Eq.~(\ref{eq:sigma-top-TOT-standard}) is known exactly through one
loop \cite{Czakon:2008cx} (see also Ref.~\cite{Hagiwara:2008df}).
Interestingly, the total cross section is the only $t{\bar t}$
observable for which at present the hard function is known beyond
the leading order with full color dependence. We have used the
degeneracy of the eigenvalues of the matrix ${\bf\Gamma}_S$ in the
gluon fusion reaction (see also appendix \ref{sec:app}) to
explicitly perform the trace over the degenerate octet eigenvalue in
Eq.~(\ref{eq:sigma-top-TOT-standard}). Therefore the hard function
${\bf  H}$ in this reaction is also a two-by-two matrix as
computed in Ref.~\cite{Czakon:2008cx}.

The
interplay of the jet
anomalous dimension $\Gamma_{\rm incl}$
with the soft matrix is
quite interesting.
As can be seen from the results in appendix
\ref{sec:app}, through two loops $\Gamma_{\rm incl}$ equals minus the singlet
component of the anomalous dimension matrix ${\bf\Gamma}_S$. Indeed,
it is natural to expect that the anomalous dimension for producing a
color singlet is the same as the one in Drell-Yan or Higgs
production, a fact that has been anticipated already in
Ref.~\cite{Bonciani:1998vc}. Reproducing this property, without
imposing it by hand, represents a very strong check on the
consistency of our setup and results.

The vanishing of the sum of the color singlet anomalous dimension
and $\Gamma_{\rm incl}$ is even more striking given the fact that
they refer to very different kinematics: the former is related to
the``very heavy mass'' limit close to partonic threshold where the
mass is comparable in value to the hard scale, while the former is
extracted from the small mass limit where the mass is negligible
compared to the hard scale. This result is quite intriguing, and it
is clear that it is not accidental, as implied by the argument that
the singlet anomalous dimension should not receive corrections
beyond Drell-Yan/Higgs. Clearly, one can explore this property to
relate the anomalous dimension in heavy flavor hadroproduction to
the small-mass limit of the form factor beyond two loops assuming,
of course, the findings of Ref.~\cite{Mitov:2009sv} extend to three
loops. Combining the results in \cite{Mitov:2006xs} and
\cite{Gluza:2009yy} one can cast the three-loop result in terms of
only one unknown constant $K_3$ (note that the function $G_3$ is
known from the massless form factor
\cite{Moch:2005id,Baikov:2009bg}).

The explicit form given in Eq.~(\ref{eq:D-QQ}) is among the main results of this work. These
anomalous dimensions provide explicit corrections to the form
proposed in Ref.~\cite{Moch:2008qy}, which was based on generalizing
the proportionality between the one- and two-loop massive anomalous
dimension matrices of the massless case \cite{Aybat}. The results of
the present work as well as of
Refs.~\cite{Mitov:2009sv,Becher:2009kw,Kidonakis:2009ev} provide the
necessary corrections, which arise even for the special kinematics
close to absolute threshold.

\section{Discussion and outlook}

In the present paper we have extended the formalism for the resummation
of soft gluon logarithms in cross sections with massive partons to the
NNLL level. A central role in our construction is played by the
massive two-loop soft anomalous dimension matrix for processes with
$n\ge 4$ colored hard partons. In this paper we have presented the most
general form of the so-called $2E$ (dipole) contributions. Combined
with the results of Ref.~\cite{Mitov:2009sv} this allows resummation
in observables with special kinematics, like the total inclusive
cross section for hadroproduction of a pair of heavy flavors.

In our discussion we have detailed the relation between the soft
function in an observable and the soft function in the
corresponding factorized (virtual) amplitude.
Following \cite{Kidonakis:1997gm}, we have shown how
the two are closely related for generic processes, and that
the infrared poles of the amplitudes can be used to
specify properties of the cross sections near partonic
threshold, particularly in the form factor scheme
defined and applied here.

We have also explained how to construct the various jet factors
needed for the completion of our threshold resummation, and have
used their process independence to derive
initial-state partonic jets from the Drell-Yan
vector production process.  We have also used heuristic arguments to
identify and construct natural jet factors needed for inclusive observables, like the
total cross section for heavy pair production.   Factorized jet functions, although
not strictly necessary, provide a form that can be extended
to cross sections at measured pair mass, even far above
absolute threshold.

The most phenomenologically relevant application of our work is the
total cross section for heavy pair production at hadron colliders.
In this paper we have shown how to derive this quantity from
resummed cross sections at fixed scattering kinematics. We have also
given the exact result for the two-loop anomalous dimensions
controlling the exponentiation of the NNLL terms in the cross
section, and we have verified that even above threshold the part of
the two-loop soft anomalous dimension constructed here is adequate
to determine exactly the pole structure of the two-loop
color-averaged amplitudes for top production through light quark
annihilation.

Our result provides not only the result
for the resummed logs to NNLL but also a framework for studying the
higher order effects in this observable and the associated
theoretical uncertainties. We will provide a detailed numerical
analysis in a dedicated publication.

The formalism we have presented here can also be applied to more
differential observables in heavy flavor production at hadron
colliders, including cross sections at fixed rapidity
and pair mass. To complete such studies one will, however, require
explicit results, whether analytic
or numerical, for the $3E$-type contributions to the anomalous dimension
matrix involving two massive partons.

\appendix\section{Color Bases}\label{sec:app}

In this appendix we present calculations of the one-loop
anomalous dimension in singlet-octet basis, and the
evaluation of the two-loop expression of Eq.\ (\ref{eq:Gamma2mass-qq-gg}).

\subsection{One-loop results in singlet-octet basis}

We work out the general result for the one-loop soft anomalous
dimension matrices in the form factor scheme for the two reactions:
\begin{eqnarray}
&& q(p_1){\bar q}(p_2) \to Q(p_3){\overline Q}(p_4)\label{eq:qq-QQ}\\
&& g(p_1)g(p_2) \to Q(p_3){\overline Q}(p_4)\label{eq:gg-QQ}
\end{eqnarray}
where $p_1^2=p_2^2=0;~p_3^2=p_4^2=m^2$. We define the invariants
$s=(p_1+p_2)^2, t_1=(p_1-p_3)^2-m^2=(p_2-p_4)^2-m^2,
u_1=(p_1-p_4)^2-m^2=(p_2-p_3)^2-m^2$. In the massive case
$\sigma_{34} = 2p_3\cdot p_4 = s-2m^2 \neq s$.

We first consider the reaction (\ref{eq:qq-QQ}). To evaluate the
color matrices $T_i\cdot T_j$ a color basis needs to be specified.
The simplest one is:
\begin{equation}
c_1 = \delta_{12}\delta_{34}~,~ c_2 = \delta_{13}\delta_{24}\, ,
\end{equation}
There are six combinations of the indices $(i,j)$ that need to be
considered. Only three of them are different. Denoting these three
color matrices by ${\bf \hat\Gamma_S},{\bf \hat\Gamma_T}$ and ${\bf
\hat\Gamma_U}$ defined through ${\bf \hat\Gamma_S}=T_1\cdot T_2$,
etc., we get:
\begin{equation}
{\bf \Gamma}_S^{(1)} = 2T~{\bf \hat\Gamma_T} + 2U~{\bf \hat\Gamma_U}
+
\left( S_0 + S_m + P\right)~{\bf \hat\Gamma_S} \nonumber\\
= \left(
  \begin{array}{cc}
   \Gamma^{(1)}_{11}  & \Gamma^{(1)}_{12} \\
   \Gamma^{(1)}_{21}  & \Gamma^{(1)}_{22} \\
  \end{array}
\right) \, , \nonumber
\end{equation}
where:
\begin{eqnarray}
\Gamma^{(1)}_{11} &=& \left(2U-2T-S_0-S_m-P\right)
C_F+\left(T-U\right) C_A \, , \nonumber\\
\Gamma^{(1)}_{12} &=& {1\over 2}\left(2U-S_0-S_m-P\right)\, , \nonumber\\
\Gamma^{(1)}_{21} &=& U-T \, , \nonumber\\
\Gamma^{(1)}_{22} &=& \left(2U-2T-S_0-S_m-P\right) C_F -
\left(2U-S_0-S_m-P\right) {C_A\over 2}  \, . \label{eq:Gamma1-c1c2}
\end{eqnarray}
The individual matrices ${\bf \hat\Gamma_S},{\bf \hat\Gamma_T}$ and
${\bf \hat\Gamma_U}$ can be easily read off from the above
equations. We have also included an overall minus sign in ${\bf
\hat\Gamma_S},{\bf \hat\Gamma_T}$ as follows from the sign
conventions for the color generators of
Ref.~\cite{Catani:1998bh,Aybat}: a generator $T_i$ is multiplied by
$-1$ if it is inserted in a line that represents incoming quark or
gluon or outgoing antiquark.

The expressions for the velocity factors $U,T,S_0,S_m$ and $P$
follow directly from Eq.~(\ref{eq:Gamma1mass-result}):
\begin{eqnarray}
&& U = \ln\left( - {\mu^2\over u_1}\right)~,~ T = \ln\left( -
{\mu^2\over t_1}\right)~,~S_0 = \ln\left( {\mu^2\over s}\right) +
i\pi~, \nonumber\\
&& S_m = \ln\left( {\mu^2\over s-2m^2}\right) + i\pi ~,~ P =
P^{(1)}_{ij}\left(\xij = -{1-\beta\over 1+\beta} + i\varepsilon\,
,\, s \geq 4m^2 \right) \, .
\end{eqnarray}
We have labeled these functions according to their respective
channel $(s,t,u)$, and as to whether they refer to a massless or massive dipole.
The function $P$ collects all power corrections in the mass, i.e. in
the massless case $S_m=S_0, P=0$.

For physical applications to heavy flavor hadroproduction one
chooses the $s$-channel singlet-octet color basis:
\begin{equation}
\label{eq:v1v8-basis} {\rm v}_1 = c_1~,~ {\rm v}_8 = -{1\over 2N}
c_1 + {1\over 2} c_2\, .
\end{equation}
The change of basis for the anomalous dimension matrix in
Eq.~(\ref{eq:Gamma1-c1c2}) (or for any one of ${\bf \hat\Gamma_S},
{\bf \hat\Gamma_T}, {\bf \hat\Gamma_U}$) is:
\begin{equation}
{\bf \Gamma}_S^{(1),{\rm S.O.}} = {\rm \bf R}^{-1} \cdot {\bf
\Gamma}_S^{(1)} \cdot {\rm \bf R} \, ,
\end{equation}
where the transformation matrix ${\rm \bf R}$ reads:
\begin{equation}
{\rm \bf R} = \left(
  \begin{array}{cc}
    1 & -{1\over 2N} \\
    0 & {1\over 2} \\
  \end{array}
\right)~~,~~ {\rm \bf R}^{-1} = \left(
  \begin{array}{cc}
    1 & {1\over N} \\
    0 & 2 \\
  \end{array}
\right) \, .
\end{equation}

The matrix elements of the matrix ${\bf \Gamma}_S^{(1),{\rm S.O.}}$
read:
\begin{eqnarray}
\Gamma^{(1),{\rm S.O.}}_{11} &=& -\left(S_0+S_m+P\right)
C_F \, , \nonumber\\
\Gamma^{(1),{\rm S.O.}}_{12} &=& \left(U-T\right) {C_F\over C_A}\, , \nonumber\\
\Gamma^{(1),{\rm S.O.}}_{21} &=& 2\left(U-T\right) \, , \nonumber\\
\Gamma^{(1),{\rm S.O.}}_{22} &=& \left(4U-4T-S_0-S_m-P\right) C_F -
\left(4U-2T-S_0-S_m-P\right) {C_A\over 2}  \, .
\label{eq:Gamma1-v1v8}
\end{eqnarray}

Rearranging the above expressions we arrive at the following
expression for the anomalous dimension matrix ${\bf
\Gamma}_S^{(1),{\rm S.O.}}$ in the singlet-octet color basis:
\begin{eqnarray}
\Gamma^{(1),{\rm S.O.}}_{11} &=& \left[ 2\ln\left({m^2\over
\mu^2}\right)
-\ln\left({m^2\over s}\right) - L_\beta- i\pi \right] C_F \, , \nonumber\\
\Gamma^{(1),{\rm S.O.}}_{12} &=& \ln\left( {t_1\over u_1} \right) {C_F\over C_A}\, , \nonumber\\
\Gamma^{(1),{\rm S.O.}}_{21} &=& 2\ln\left( {t_1\over u_1} \right) \, , \nonumber\\
\Gamma^{(1),{\rm S.O.}}_{22} &=& \left[ 4\ln\left( {t_1\over u_1}
\right) + 2\ln\left({m^2\over \mu^2}\right)-\ln\left({m^2\over
s}\right) -L_\beta-i\pi \right] C_F \nonumber\\
&-& \left[ 2\ln\left( {t_1\over u_1} \right) - 2\ln\left( -{u_1\over
s} \right) + \ln\left({m^2\over s}\right) -L_\beta-i\pi \right]
{C_A\over 2} \, .\label{eq:Gamma1-qqQQ-explicit-result}
\end{eqnarray}
We have kept the traditional notations and defined:
\begin{equation}
L_\beta = {1+\beta^2\over 2\beta} \left( \ln\left({1-\beta\over
1+\beta}\right) + i\pi\right) \, .
\end{equation}

The result for the anomalous dimension matrix in ${\bf
\Gamma}_S^{(1),{\rm S.O.}}$ in
Eq.~(\ref{eq:Gamma1-qqQQ-explicit-result}) agrees with the one
derived first in Ref.~\cite{Kidonakis:1997gm} provided we add to the
above result (\ref{eq:Gamma1-qqQQ-explicit-result}) the term
$(-\ln(m^2/\mu^2)-1)C_F {\bf 1}$ and set $\mu^2=s$. The addition of
this (color diagonal) term corresponds to working in a scheme where
the two jet factors for the heavy quark and antiquark are absorbed
into the soft function (see the discussion in section
\ref{sec:NNLL-res-crossection}).

Finally, we give the expression for the real part of the anomalous
dimension matrix ${\bf \Gamma}_S^{(1),{\rm S.O.}}$ at absolute
threshold. To that end we set $u_1=t_1=-s/2$ as well as
$\ln(\mu^2/s)=0$. Up to corrections $\sim {\cal O}(\beta)$ the
result reads:
\begin{eqnarray}
{\rm Re}\, {\bf\Gamma}_S^{(1),{\rm S.O.}} &=& C_F\, \left[
\ln\left({m^2\over \mu^2}\right) + 1 \right] \left(
  \begin{array}{cc}
   1  & 0 \\
   0  & 1 \\
  \end{array}
\right) -{C_A \over 2} \left(
  \begin{array}{cc}
   0  & 0 \\
   0  & 1 \\
  \end{array}
\right) \nonumber\\
&=& -\Gamma_{\rm incl}^{(1)} \left(
  \begin{array}{cc}
   1  & 0 \\
   0  & 1 \\
  \end{array}
\right) -{C_A \over 2} \left(
  \begin{array}{cc}
   0  & 0 \\
   0  & 1 \\
  \end{array}
\right) \, . \label{eq:Gamma1-qqQQ-at-threshold}
\end{eqnarray}

\subsection{Evaluation of Eq.\ (\ref{eq:Gamma2mass-qq-gg})}

The threshold limit of the two-loop matrix ${\bf \Gamma}_S^{(2)}$ in
the basis (\ref{eq:v1v8-basis}) can be easily derived from
Eq.~(\ref{eq:Gamma2mass-qq-gg}) by using that
\begin{equation}
\label{eq:T3T4} T_3\cdot T_4 = {\bf \hat\Gamma_S^{\rm \bf S.O.}} =
\left(
  \begin{array}{cc}
   -C_F  & 0 \\
   0  & -C_F +{C_A\over 2} \\
  \end{array}
\right) \, .
\end{equation}
Up to corrections $\sim {\cal O}(\beta)$, the result for ${\bf
\Gamma}_S^{(2)}$ reads:
\begin{eqnarray}
\label{eq:Gamma2-qqQQ-at-threshold} {\rm Re}\,
{\bf\Gamma}_S^{(2),{\rm S.O.}} = -\Gamma_{\rm incl}^{(2)} \left(
  \begin{array}{cc}
   1  & 0 \\
   0  & 1 \\
  \end{array}
\right) -\left({C_A K\over 4}+{\zeta_3-1\over 4}C_A^2\right) \left(
  \begin{array}{cc}
   0  & 0 \\
   0  & 1 \\
  \end{array}
\right) \, ,
\end{eqnarray}
where $\Gamma_{\rm incl}^{(1)}$ and $\Gamma_{\rm incl}^{(2)}$ are
given in Eq.~(\ref{eq:Gamma-inclusive}).

The calculation of the gluon fusion reaction Eq.~(\ref{eq:gg-QQ})
can be done in the same way as in the quark-antiquark annihilation
one described above. The appropriate $s$-channel singlet-octet color
basis is:
\begin{equation}
{\rm v}_1 = \delta_{12}\delta_{34} ~,~ {\rm v}_8^s = d^{12c}T^c_{34}
~,~ {\rm v}_8^a = if^{12c}T^c_{34} \, .
\end{equation}
A direct calculation shows that in the limit $\beta \to 0$ the
matrices $T_3\cdot T_4,{\rm Re}\, {\bf\Gamma}_S^{(1),{\rm S.O.}}$
and ${\rm Re}\, {\bf\Gamma}_S^{(2),{\rm S.O.}}$ for the gluon fusion
reaction can be obtained from the corresponding matrices in the
quark-antiquark initiated one
Eqs.~(\ref{eq:T3T4},\ref{eq:Gamma1-qqQQ-at-threshold},\ref{eq:Gamma2-qqQQ-at-threshold})
with the help of the simple replacements:
\begin{eqnarray}
\label{eq:replacement} {\bf 1}_{\bf 2\times 2}\longrightarrow {\bf
1}_{\bf 3\times 3} ~~~ {\rm and}~~~ \left(
  \begin{array}{cc}
   0  & 0 \\
   0  & 1 \\
  \end{array}
\right) \longrightarrow \left(
  \begin{array}{ccc}
   0  & 0 & 0\\
   0  & 1 & 0\\
   0  & 0 & 1\\
  \end{array}
\right)\, .
\end{eqnarray}
The results of Ref.~\cite{Kidonakis:1998nf} can be used to simplify
the calculations.

\acknowledgments We would like to thank A.~Smirnov and M.~Tentyukov
for their kind help with the program FIESTA \cite{Smirnov:2008py} as
well as R.~Bonciani and A.~Ferroglia for useful communication
regarding Ref.~\cite{Bonciani:2008wf}. We also thank J.~Gluza and
T.~Riemann for related collaboration, and L.\ Almeida, I.\ Sung
and W.\ Vogelsang for helpful conversations. This work was supported in part by the
National Science Foundation, grants PHY-0354776, PHY-0354822 and
PHY-0653342. The work of M.C. was supported by the Heisenberg
Programme of the Deutsche Forschungsgemeinschaft.  The work of A.M. is
supported by a fellowship from the {\it US LHC Theory Initiative}
through NSF grant 0705682.

While this work was being finalized, we became aware of independent
research on the subject \cite{Beneke:2009rj}. We would like to thank
M. Beneke for communicating the results of that study which relate to
the NNLL resummation shown in Eq.\ (\ref{eq:sigma-top-TOT})-(\ref{eq:D-QQ}) above before
publication. A comparison showed full agreement.


\begin{thebibliography}{}


\bibitem{QQbarfixed}
  P.~Nason, S.~Dawson and R.~K.~Ellis,
  Nucl.\ Phys.\  B {\bf 303}, 607 (1988);
  W.~Beenakker, H.~Kuijf, W.~L.~van Neerven and J.~Smith,
  Phys.\ Rev.\  D {\bf 40}, 54 (1989);

\bibitem{Czakon:2008ii}
  M.~Czakon and A.~Mitov,
  arXiv:0811.4119 [hep-ph].

  \bibitem{dyresum} G.~Sterman, Nucl.\ Phys.\ B {\bf 281}, 310 (1987);
S.~Catani and L.~Trentadue, Nucl.\ Phys.\ B {\bf 327}, 323 (1989);
Nucl.\ Phys.\ B {\bf 353}, 183 (1991).

\bibitem{Bonciani:1998vc}
  R.~Bonciani, S.~Catani, M.~L.~Mangano and P.~Nason,
  Nucl.\ Phys.\  B {\bf 529}, 424 (1998)
  [Erratum-ibid.\  B {\bf 803}, 234 (2008)]
  [arXiv:hep-ph/9801375];

\bibitem{Kidonakis:1997gm}
N.~Kidonakis and G.~Sterman,
Nucl.\ Phys.\  B {\bf 505}, 321 (1997) [arXiv:hep-ph/9705234];
N.~Kidonakis, G.~Oderda and G.~Sterman, Nucl. Phys. {\bf B531}
(1998) 365 [hep-ph/9803241].

\bibitem{Mitov:2009sv}
  A.~Mitov, G.~Sterman and I.~Sung,
  Phys.\ Rev.\  D {\bf 79}, 094015 (2009)
  [arXiv:0903.3241 [hep-ph]].

\bibitem{Kidonakis:2009ev}
  N.~Kidonakis,
  Phys.\ Rev.\ Lett.\  {\bf 102}, 232003 (2009)
  [arXiv:0903.2561 [hep-ph]].

\bibitem{Becher:2009kw}
  T.~Becher and M.~Neubert,
  Phys.\ Rev.\  D {\bf 79}, 125004 (2009)
  [arXiv:0904.1021 [hep-ph]].

\bibitem{Contopanagos:1996nh}
  H.~Contopanagos, E.~Laenen and G.~Sterman,
  Nucl.\ Phys.\  B {\bf 484}, 303 (1997)
  [arXiv:hep-ph/9604313].

\bibitem{Kidonakis:2001nj}
  N.~Kidonakis, E.~Laenen, S.~Moch and R.~Vogt,
  Phys.\ Rev.\  D {\bf 64}, 114001 (2001)
  [arXiv:hep-ph/0105041].

\bibitem{Sen:1982bt}
  A.~Sen,
  Phys.\ Rev.\  D {\bf 28}, 860 (1983).

\bibitem{catani96}
S.~Catani and M.~H.~Seymour,
Phys.\ Lett.\ {\bf B378}, 287 (1996)
[hep-ph/9602277];
Nucl.\ Phys.\ {\bf B485}, 291 (1997)
[Err.-ibid.\ {\bf B510}, 503 (1997)]
[hep-ph/9605323].

\bibitem{Aybat} 
  S.~Mert Aybat, L.~J.~Dixon and G.~Sterman,
  Phys.\ Rev.\ Lett.\  {\bf 97}, 072001 (2006)
  [arXiv:hep-ph/0606254];
  S.~Mert Aybat, L.~J.~Dixon and G.~Sterman,
  Phys.\ Rev.\  D {\bf 74}, 074004 (2006)
  [arXiv:hep-ph/0607309].

\bibitem{Becher:2009cu}
  T.~Becher and M.~Neubert,
  Phys.\ Rev.\ Lett.\  {\bf 102} (2009) 162001
  [arXiv:0901.0722 [hep-ph]].

\bibitem{Gardi:2009qi}
  E.~Gardi and L.~Magnea,
  JHEP {\bf 0903}, 079 (2009)
  [arXiv:0901.1091 [hep-ph]].

\bibitem{Becher:2009qa}
  T.~Becher and M.~Neubert,
  arXiv:0903.1126 [hep-ph].

\bibitem{TYS}
G.~Sterman and M.~E.~Tejeda-Yeomans,
Phys. Lett. {\bf B552} (2003) 48 [hep-ph/0210130].

\bibitem{Mitov:2006xs}
A.~Mitov and S.~Moch,
JHEP {\bf 0705}, 001 (2007) [arXiv:hep-ph/0612149].

\bibitem{Magnea:1990zb}
  L.~Magnea and G.~Sterman,
  Phys.\ Rev.\  D {\bf 42}, 4222 (1990).

\bibitem{Magnea:2000ss}
  L.~Magnea,
  Nucl.\ Phys.\  B {\bf 593}, 269 (2001)
  [arXiv:hep-ph/0006255].

\bibitem{Moch:2005id}
  S.~Moch, J.~A.~M.~Vermaseren and A.~Vogt,
  JHEP {\bf 0508}, 049 (2005)
  [arXiv:hep-ph/0507039].

\bibitem{Moch:2005ky}
  S.~Moch and A.~Vogt,
  Phys.\ Lett.\  B {\bf 631}, 48 (2005)
  [arXiv:hep-ph/0508265].

\bibitem{Moch:2005ba}
  S.~Moch, J.~A.~M.~Vermaseren and A.~Vogt,
  Nucl.\ Phys.\  B {\bf 726}, 317 (2005)
  [arXiv:hep-ph/0506288].

\bibitem{Moch:2004pa}
  S.~Moch, J.~A.~M.~Vermaseren and A.~Vogt,
  Nucl.\ Phys.\  B {\bf 688}, 101 (2004)
  [arXiv:hep-ph/0403192].

\bibitem{Vogt:2004mw}
  A.~Vogt, S.~Moch and J.~A.~M.~Vermaseren,
  Nucl.\ Phys.\  B {\bf 691}, 129 (2004)
  [arXiv:hep-ph/0404111].

\bibitem{Laenen:2005uz}
  E.~Laenen and L.~Magnea,
  Phys.\ Lett.\  B {\bf 632}, 270 (2006)
  [arXiv:hep-ph/0508284].

\bibitem{Becher:2007ty}
  T.~Becher, M.~Neubert and G.~Xu,
  JHEP {\bf 0807}, 030 (2008)
  [arXiv:0710.0680 [hep-ph]].


\bibitem{Gardi:2005yi}
  E.~Gardi,
  JHEP {\bf 0502}, 053 (2005)
  [arXiv:hep-ph/0501257].

\bibitem{Mele:1990cw}
  B.~Mele and P.~Nason,
  Nucl.\ Phys.\  B {\bf 361}, 626 (1991).

\bibitem{Melnikov:2004bm}
  K.~Melnikov and A.~Mitov,
  Phys.\ Rev.\  D {\bf 70}, 034027 (2004)
  [arXiv:hep-ph/0404143].

\bibitem{Cacciari:2001cw}
  M.~Cacciari and S.~Catani,
  Nucl.\ Phys.\  B {\bf 617}, 253 (2001)
  [arXiv:hep-ph/0107138].

\bibitem{Rijken:1996ns}
  P.~J.~Rijken and W.~L.~van Neerven,
  Nucl.\ Phys.\  B {\bf 487}, 233 (1997)
  [arXiv:hep-ph/9609377].

\bibitem{Mitov:2006wy}
  A.~Mitov and S.~O.~Moch,
  Nucl.\ Phys.\  B {\bf 751}, 18 (2006)
  [arXiv:hep-ph/0604160].

\bibitem{Sterman:2006hu}
  G.~Sterman and W.~Vogelsang,
  Phys.\ Rev.\  D {\bf 74}, 114002 (2006)
  [arXiv:hep-ph/0606211].

\bibitem{Beneke:2009rj}
  M.~Beneke, P.~Falgari and C.~Schwinn,
  arXiv:0907.1443 [hep-ph].

\bibitem{Ferroglia:2009ep}
  A.~Ferroglia, M.~Neubert, B.~D.~Pecjak and L.~L.~Yang,
  arXiv:0907.4791 [hep-ph];
  A.~Ferroglia, M.~Neubert, B.~D.~Pecjak and L.~L.~Yang,
  arXiv:0908.3676 [hep-ph].

\bibitem{CMSI-prep}
  M.~Czakon, A.~Mitov,  G.~Sterman and I.~Sung, in preparation.

\bibitem{Catani:2000ef}
  S.~Catani, S.~Dittmaier and Z.~Trocsanyi,
  Phys.\ Lett.\  B {\bf 500}, 149 (2001)
  [arXiv:hep-ph/0011222].

\bibitem{Bernreuther:2004ih}
  W.~Bernreuther, R.~Bonciani, T.~Gehrmann, R.~Heinesch, T.~Leineweber, P.~Mastrolia and E.~Remiddi,
  Nucl.\ Phys.\  B {\bf 706}, 245 (2005)
  [arXiv:hep-ph/0406046].

\bibitem{Bonciani:2008wf}
  R.~Bonciani and A.~Ferroglia,
  JHEP {\bf 0811}, 065 (2008)
  [arXiv:0809.4687 [hep-ph]].

\bibitem{Beneke:2008ei}
  M.~Beneke, T.~Huber and X.~Q.~Li,
  Nucl.\ Phys.\  B {\bf 811}, 77 (2009)
  [arXiv:0810.1230 [hep-ph]].

\bibitem{Bell:2006tz}
  G.~Bell,
  arXiv:0705.3133 [hep-ph].

\bibitem{Asatrian:2008uk}
  H.~M.~Asatrian, C.~Greub and B.~D.~Pecjak,
  Phys.\ Rev.\  D {\bf 78}, 114028 (2008)
  [arXiv:0810.0987 [hep-ph]].

\bibitem{Huber:2009se}
  T.~Huber,
  JHEP {\bf 0903}, 024 (2009)
  [arXiv:0901.2133 [hep-ph]].

\bibitem{HPL}
  D.~Maitre,
  Comput.\ Phys.\ Commun.\  {\bf 174}, 222 (2006)
  [arXiv:hep-ph/0507152];
  D.~Maitre,
  arXiv:hep-ph/0703052.

\bibitem{Smirnov:2008py}
  A.~V.~Smirnov and M.~N.~Tentyukov,
  Comput.\ Phys.\ Commun.\  {\bf 180}, 735 (2009)
  [arXiv:0807.4129 [hep-ph]].

\bibitem{Gluza:2009yy}
  J.~Gluza, A.~Mitov, S.~Moch and T.~Riemann,
  arXiv:0905.1137 [hep-ph].

\bibitem{Korchemsky:1987wg}
  G.~P.~Korchemsky and A.~V.~Radyushkin,
  Nucl.\ Phys.\  B {\bf 283}, 342 (1987);
  G.~P.~Korchemsky and A.~V.~Radyushkin,
  Phys.\ Lett.\  B {\bf 279}, 359 (1992)
  [arXiv:hep-ph/9203222].

\bibitem{Czakon:2008zk}
  M.~Czakon,
  Phys.\ Lett.\  B {\bf 664}, 307 (2008)
  [arXiv:0803.1400 [hep-ph]].

\bibitem{Czakon:2007ej}
  M.~Czakon, A.~Mitov and S.~Moch,
  Phys.\ Lett.\  B {\bf 651}, 147 (2007)
  [arXiv:0705.1975 [hep-ph]];
  M.~Czakon, A.~Mitov and S.~Moch,
  Nucl.\ Phys.\  B {\bf 798}, 210 (2008)
  [arXiv:0707.4139 [hep-ph]].

\bibitem{Bonciani:2008az}
  R.~Bonciani, A.~Ferroglia, T.~Gehrmann, D.~Maitre and C.~Studerus,
  JHEP {\bf 0807}, 129 (2008)
  [arXiv:0806.2301 [hep-ph]].
  R.~Bonciani, A.~Ferroglia, T.~Gehrmann and C.~Studerus,
  arXiv:0906.3671 [hep-ph].

\bibitem{Dixon:2008gr}
  L.~J.~Dixon, L.~Magnea and G.~Sterman,
  JHEP {\bf 0808}, 022 (2008)
  [arXiv:0805.3515 [hep-ph]].

\bibitem{Catani:1999ss}
  S.~Catani and M.~Grazzini,
  Nucl.\ Phys.\  B {\bf 570}, 287 (2000)
  [arXiv:hep-ph/9908523].


\bibitem{Czakon:2008cx}
  M.~Czakon and A.~Mitov,
  arXiv:0812.0353 [hep-ph].

\bibitem{Hagiwara:2008df}
  K.~Hagiwara, Y.~Sumino and H.~Yokoya,
  Phys.\ Lett.\  B {\bf 666}, 71 (2008)
  [arXiv:0804.1014 [hep-ph]].

\bibitem{Baikov:2009bg}
  P.~A.~Baikov, K.~G.~Chetyrkin, A.~V.~Smirnov, V.~A.~Smirnov and M.~Steinhauser,
  Phys.\ Rev.\ Lett.\  {\bf 102}, 212002 (2009)
  [arXiv:0902.3519 [hep-ph]].

\bibitem{Moch:2008qy}
  S.~Moch and P.~Uwer,
  Phys.\ Rev.\  D {\bf 78}, 034003 (2008)
  [arXiv:0804.1476 [hep-ph]].

\bibitem{Catani:1998bh}
  S.~Catani,
  Phys.\ Lett.\  B {\bf 427}, 161 (1998)
  [arXiv:hep-ph/9802439].

\bibitem{Kidonakis:1998nf}
  N.~Kidonakis, G.~Oderda and G.~Sterman,
  Nucl.\ Phys.\  B {\bf 531}, 365 (1998)
  [arXiv:hep-ph/9803241].Mo



\end{thebibliography}
\end{document}